\documentclass[10pt,journal]{IEEEtran}

\usepackage{amsmath,epsfig,amssymb,verbatim}
\usepackage{cite}
\usepackage{graphicx}
\usepackage{bm,bbm}
\usepackage{xcolor}
\usepackage{hhline}
\usepackage{subfigure}
\usepackage{stackengine}
\usepackage{cleveref}
\usepackage{multicol}
\usepackage{multirow}
\def\delequal{\mathrel{\ensurestackMath{\stackon[1pt]{=}{\scriptstyle\Delta}}}}

\def\@IEEEBIOskipN{1\baselineskip}
\newcommand{\E}{\mathbb{E}}

\newtheorem{Theorem}{Theorem}
\newtheorem{Lemma}{Lemma}

\begin{document}
\title{Impact of Pointing Errors and Correlated Wall Blockages on Practical Grid-based Indoor Terahertz Communication Systems}

\author{Zhifeng Tang, \IEEEmembership{Member, IEEE,} Nan Yang, \IEEEmembership{Senior Member, IEEE,} Salman Durrani, \IEEEmembership{Senior Member, IEEE,} Xiangyun Zhou, \IEEEmembership{Fellow, IEEE,} Josep Miquel Jornet, \IEEEmembership{Fellow, IEEE,} and Markku Juntti, \IEEEmembership{Fellow, IEEE}

\thanks{This work was supported by the Australian Research Council Discovery Project (DP230100878). The work by J. M. Jornet was supported in part by the National Science Foundation (NSF) grant CNS-2225590. The work by M. Juntti was supported in part by the Research Council of Finland by 6G Flagship (grant 369116).}
\thanks{Z. Tang, N. Yang, S. Durrani, and X. Zhou are with the School of Engineering, Australian National University, Canberra, ACT 2601, Australia (Emails: \{zhifeng.tang, nan.yang, salman.durrani, xiangyun.zhou\}@anu.edu.au).}
\thanks{J. M. Jornet is with the Department of Electrical and Computer Engineering, Northeastern University, Boston, MA 02120, USA (Email: j.jornet@northeastern.edu).}
\thanks{M. Juntti is with the Centre for Wireless Communications, University of Oulu, Oulu 90014, Finland (Email: markku.juntti@oulu.fi).}}

\maketitle

\begin{abstract}
Terahertz (THz) communications has emerged as a promising technology for future wireless systems due to its potential to support extremely high data rates. However, severe path loss, blockage effects, and sensitivity to beam misalignment pose major challenges to reliable indoor THz communications. In this paper, we investigate the coverage probability of downlink transmission in a three-dimensional (3D) indoor THz communication system under structured access point (AP) deployments, with a focus on square and hexagonal grid topologies. A tractable analytical framework is developed to jointly account for human blockages, correlated wall blockages across APs, beam training, and residual pointing error. Numerical results demonstrate that wall blockage correlation significantly reduces the association and coverage probabilities, and its impact cannot be neglected in system performance analysis. Compared with square grid AP deployments, hexagonal grids consistently achieve higher coverage by mitigating correlated wall blockage effects and reducing the distances between user equipments (UEs) and their associated APs. Furthermore, coverage performance is shown to strongly depend on the UE location, with noticeable degradation as the UE moves away from its nearest AP. Residual pointing error is found to introduce substantial coverage loss, especially for longer links. In addition, beam training analysis reveals a non-monotonic relationship between antenna array size and training overhead, highlighting an inherent tradeoff among antenna configuration, beamwidth selection, and beam training efficiency. These findings provide useful insights into the design and deployment of practical indoor THz communication systems.
\end{abstract}
\vspace{-1.5em}


\vspace{-3mm}

\section{Introduction}

Terahertz (THz) communications has attracted significant research interest as a promising approach for future high-capacity wireless systems, due to its potential to enable extremely high data rates and dense user connectivity \cite{6GNet2020}. The ETSI ISG THz group is currently doing the pre-work for possible future inclusion of THz wireless systems in six generation (6G) standardization activities which are expected to start in 2027 \cite{Zugno2025wcm}. In particular, THz communications operate in the frequency range of 0.1 to 10~THz, providing access to extensive spectral resources \cite{THzComMCS2024,Elayan2020ojcom}. These spectral characteristics make THz communications especially attractive for short-range applications that demand high data rates \cite{Shafie2023,tang2025tvt}. However, operation at such high frequencies also introduces severe propagation impairments. Specifically, THz links experience pronounced path loss, molecular absorption, and strong sensitivity to blockage and beam misalignment, which fundamentally limit both coverage and reliable signal propagation \cite{Jornet2011,Han2016tsp}. These factors indicate that the performance of THz communication systems is highly sensitive to the propagation environment.

Motivated by the unique propagation characteristics of THz communications, indoor environments are widely regarded as one of the most promising application scenarios due to its short transmission distances and relatively controlled propagation conditions \cite{Akyildiz2018Cma}. Therefore, evaluating system performance in realistic indoor THz communication systems has attracted significant research interest. Existing studies have extensively investigated the coverage performance of THz systems by leveraging stochastic geometry frameworks and incorporating highly directional antennas to compensate for severe spreading loss and molecular absorption \cite{Yu2017JSAC,Rebato2019tcom}. In this context, the coverage probability of THz systems has been analyzed by accounting for interference effects \cite{Kokkoniemi2017twc,Petrov2017} as well as indoor propagation characteristics, including line-of-sight (LoS) and non-line-of-sight (NLoS) signal transmission \cite{Yao2019iccc}. Moreover, the impact of blockages caused by humans and walls has been investigated in both two-dimensional (2D) and three-dimensional (3D) indoor THz systems, highlighting their critical influence on coverage performance \cite{Venugopal2016Access,Wu2021TWC,Shafie2021JSAC}. Recent works have further extended these studies by considering joint blockage effects, hybrid sub-6 gigahertz (GHz) and THz architectures, and refined channel models to more accurately characterize indoor THz propagation \cite{Kouzayha2023twc,Tang2025Tcom}. However, the aforementioned studies generally rely on stochastic geometry–based spatial models to characterize the locations of access points (APs), which may not accurately reflect practical indoor deployment layouts. In realistic indoor environments for THz use cases, such as a home or office or classroom and medical facility or factory \cite{Zugno2025wcm}, APs are typically expected to be installed following structured layouts, and neglecting such deployment regularity can lead to inaccurate characterization of user–AP distances and blockage correlations, which in turn causes inaccurate interference modeling and consequently leads to ineffective coverage performance evaluation. Therefore, it is important to investigate THz system performance under more practical and structured AP deployment topologies.

In addition to deployment topology and propagation effects, the highly directional nature of THz communications introduces further challenges in practical system operation. To compensate for severe path loss, THz systems rely on narrow beams formed by large antenna arrays, which requires accurate beam alignment between transmitters and receivers \cite{xu2025empiricalstudynearfieldspatial}. In practice, such alignment is typically established through beam training procedures \cite{Chen2024twc,Zheng2025twc,Chen2025twc}. However, due to finite training overhead, estimation inaccuracies, hardware impairments, and user mobility, perfect beam alignment is difficult to achieve \cite{Zheng2025jsac}. Moreover, even with correctly performed beam training, residual pointing errors may still exist due to finite beamwidth, angular quantization, and practical alignment imperfections. In addition, as the number of antenna elements increases, the beam training overhead grows significantly owing to the expanded beam search space, which introduces a critical tradeoff between alignment accuracy and training efficiency \cite{Ning2022twc}. As a consequence, the pointing error inevitably arises and can significantly degrade the received signal power and overall system performance \cite{Boulogeorgos2022,Petrov2020tvt}. Motivated by these challenges, several studies have investigated the impact of pointing error on THz communication systems. For instance, \cite{Dabiri2022wcl} developed an analytical framework to characterize pointing error in THz links, providing fundamental insights into its severity. Building on this, \cite{Safahan2025coml} proposed an analytical pointing error model for highly directional THz transmissions, enabling tractable performance evaluation under practical conditions. In addition, \cite{Dabiri2023ojcom} examined the influence of pointing error on the outage probability in unmanned aerial vehicle (UAV)-assisted THz communication systems and proposed optimization strategies to mitigate mobility-induced misalignment. More recently, \cite{tang2026impactpointingerrorcoverage} investigated the impact of pointing error on the coverage probability of indoor THz communication systems, demonstrating that coverage performance is fundamentally constrained by beam misalignment. While these studies provide valuable insights into the modeling and impact of pointing error, its effect on the coverage performance of indoor THz communication systems, particularly when considering practical beam training processes, has not been adequately investigated, which motivates this work.

\textit{Our Contributions:} 
To address the above challenges, this paper develops an integrated analytical framework that connects network geometry, propagation characteristics, and beam training mechanisms, recognizing that indoor THz system performance is jointly determined by deployment topology, blockage correlation, and beam-level operation. In particular, we investigate the coverage probability of downlink transmission in a 3D indoor THz communication system by jointly accounting for practical deployment topology, blockage effects, beam training, and pointing error. The main contributions of this work are summarized as follows.

\begin{itemize}
    \item We develop a comprehensive analytical framework for evaluating the coverage probability of 3D indoor THz communication systems under practical structured AP deployments, including square and hexagonal grid topologies. The framework explicitly captures the impact of deployment geometry on distances between user equipments (UEs) and APs, thereby revealing coverage behaviors that fundamentally differ from those predicted by stochastic geometry models with randomly deployed APs.

    \item We incorporate a realistic wall blockage model based on Manhattan Poisson line processes (MPLPs) and explicitly characterize the resulting blockage correlation across multiple AP-UE links. The analysis demonstrates that neglecting such correlation leads to systematic overestimation of both association and coverage probabilities, particularly in densely deployed indoor environments with frequent wall blockages.

    \item We explicitly integrate beam training and residual pointing error into the system-level coverage analysis. By modeling the beam misalignment induced by finite-resolution beam training, we derive an approximate yet accurate closed-form expression for the probability distribution function (PDF) and cumulative distribution function (CDF) of the pointing error loss. We show that residual pointing error imposes an intrinsic limitation on achievable coverage when ultra-narrow beams are employed in indoor THz systems.

    \item Based on the derived analytical expressions, extensive numerical evaluations are conducted to investigate the interplay among deployment topology, beamwidth, pointing error, and antenna size. The results demonstrate that hexagonal grid AP deployment consistently outperforms square grid AP deployment by mitigating correlated blockage effects and reducing UE-AP distances. Moreover, we reveal a non-monotonic relationship between beamwidth-related antenna characteristics and beam training overhead, providing important design insights for practical indoor THz networks.

\end{itemize}

The rest of the paper is organized as follows. Section~\ref{Sec:System} presents the system model, antenna configuration, and THz channel model. Section~\ref{Sec:Coverage} analyzes the impact of pointing error and wall blockages on the received signal, AP association, and interference, followed by the derivation of the coverage probability and an investigation of beam training performance. Numerical and simulation results are presented in Section~\ref{Sec:Num}, and the paper is concluded in Section \ref{Sec:Conclusion}.

\begin{figure*}[ht]
    \centering
    \subfigure[The 3D perspective view of the considered system.]{\label{fig:VerticalView}
             \includegraphics[width=0.33\textwidth]{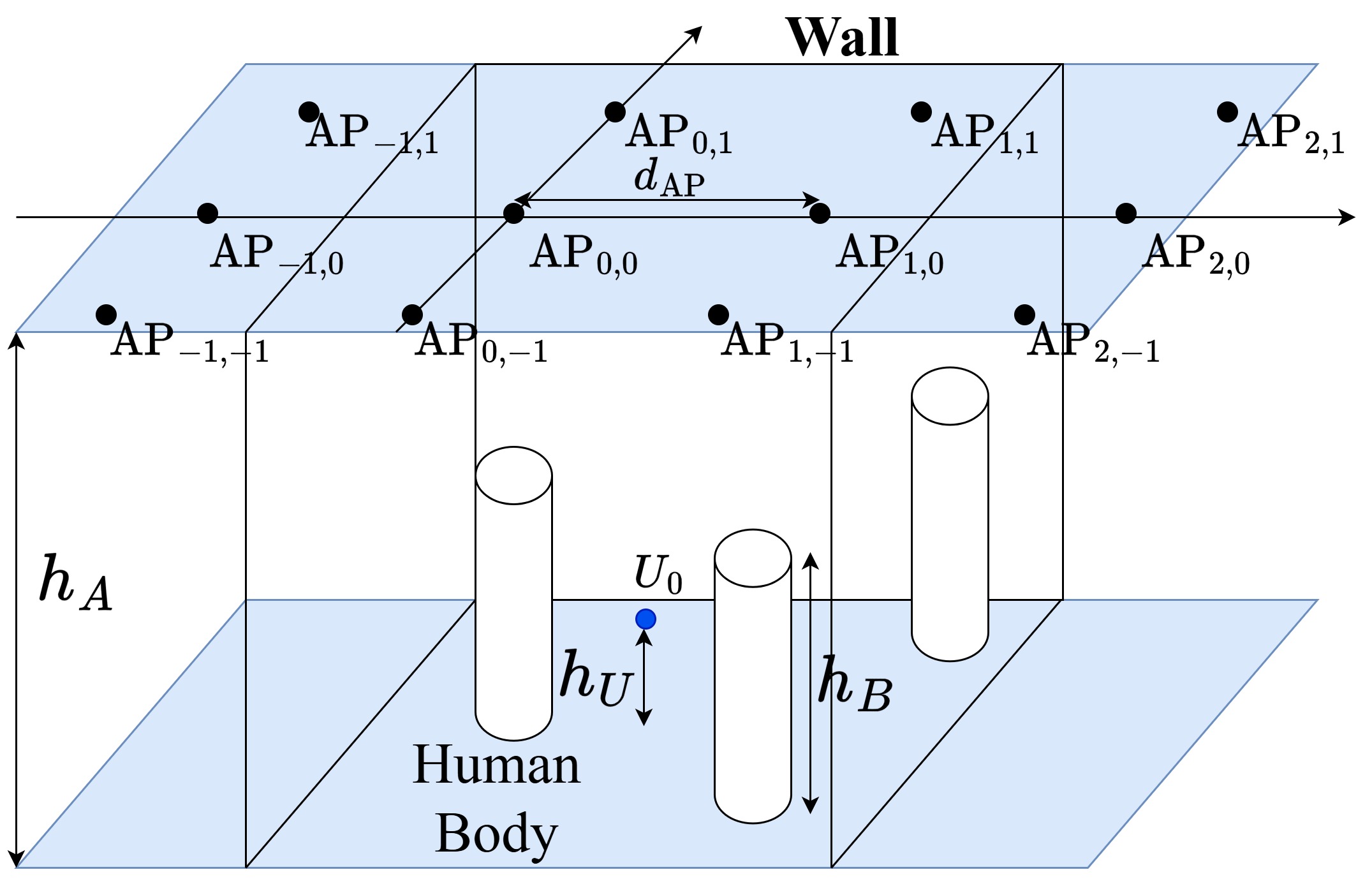}
            }
        \subfigure[The top view of the square grid.]{\label{fig:TopviewS}
             \includegraphics[width=0.3\textwidth]{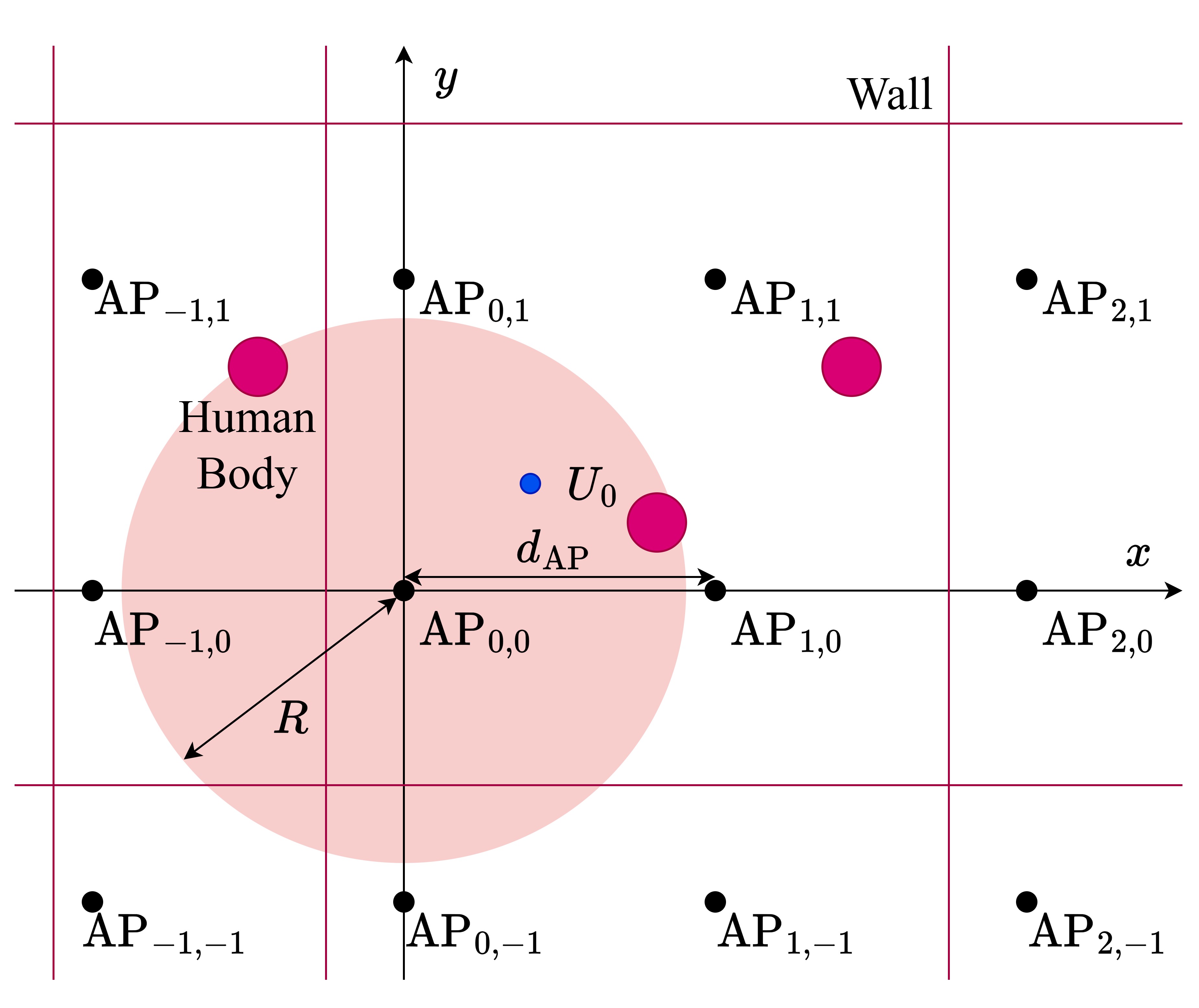}
            }
        \subfigure[The top view of the hexagonal grid.]{    \label{fig:TopviewH}        
            \includegraphics[width=0.3\textwidth]{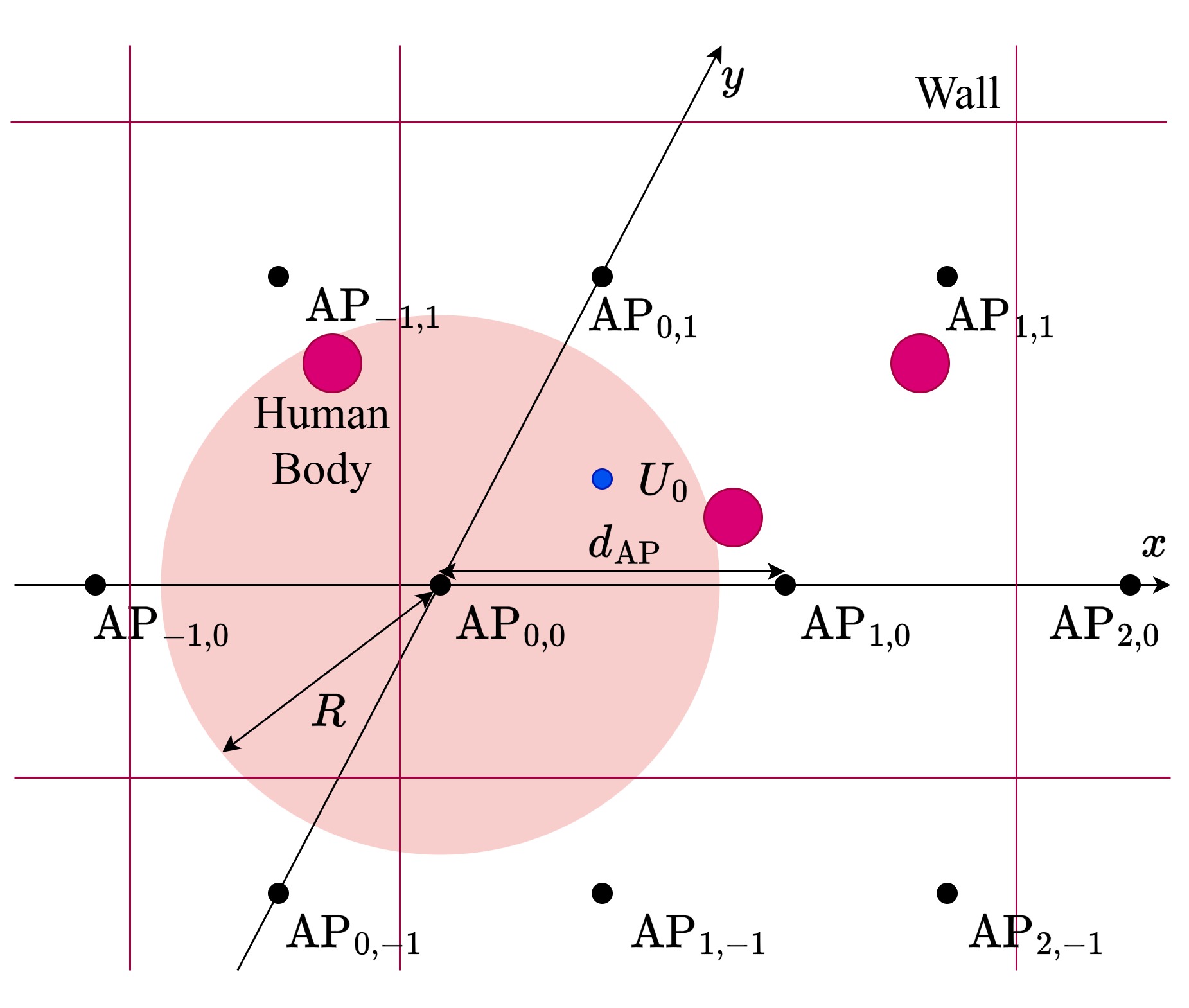}
            }
       \vspace{-0.5em}
        \caption{Illustration of the considered 3D indoor THz communication system with structured AP deployments, including square and hexagonal grid topologies.}
        \label{fig:APdeployment}\vspace{-1em}
\end{figure*}

\section{System Model}\label{Sec:System}

We consider a generalized 3D indoor THz communication system, as illustrated in Fig.~\ref{fig:APdeployment}, where multiple ceiling-mounted APs transmit THz signals to on-the-ground UEs. Specifically, the ceiling is assumed to have a fixed height $h_A$, while UEs, with a fixed height $h_U$, are randomly distributed. We consider two AP deployment strategies: (i) a square grid and (ii) a hexagonal grid, with an inter-AP distance denoted by $d_{\mathrm{AP}}$. We assume that each AP covers a circular area centered at its location, with a horizontal coverage boundary defined by radius $R_A$, and only UEs located within this region are eligible for association. In this system, one UE is randomly selected and referred to as the typical UE, denoted by $U_0$.

\subsection{System Deployment}

For analytical convenience, we establish a horizontal coordinate system on the ceiling plane with the nearest AP to the typical UE $U_0$ chosen as the origin $(0,0)$. The x-axis and y-axis are aligned with the directions from the nearest AP to the second and third nearest APs, respectively. The coordinates of all other APs are specified relative to this origin. Each AP is indexed by integer coordinates $(i,j)$ and denoted by $\mathrm{AP}_{i,j}$, corresponding to its position in the grid AP deployment. The set of all APs is denoted as $\Psi_{AP} = \{\mathrm{AP}_{i,j}|i,j\in \mathbbm{Z}\}$. Moreover, the horizontal coordinate of the typical UE is denoted by ${\mathcal{O}_{U_0}}=(x_0,y_0)$, normalized with respect to the inter-AP distance $d_{\mathrm{AP}}$. The actual physical positions of the APs and $U_0$ are expressed according to the two AP deployment strategies as follows:

\begin{itemize}
    \item \textbf{Square Grid:} In the square grid AP deployment, the x- and y-axes are orthogonal by construction, as illustrated in Fig.~\ref{fig:TopviewS}. The actual physical positions of $\mathrm{AP}_{i,j}$ and $U_0$ are obtained by scaling the grid coordinates by the inter-AP distance $d_{\mathrm{AP}}$, i.e., $\mathbf{L}_{i,j} = (i d_{\mathrm{AP}}, j d_{\mathrm{AP}})$ and $\mathbf{L}_{U_0} = (x_0 d_{\mathrm{AP}}, y_0 d_{\mathrm{AP}})$, where $0 \leq y_0 \leq x_0 \leq \frac{1}{2}$.

    \item \textbf{Hexagonal Grid:} In the hexagonal grid AP deployment, the x- and y-axes are non-orthogonal. The x-axis is aligned with the direction from the nearest AP to the second nearest AP, while the y-axis follows the direction from the nearest AP to the third nearest AP, forming a $60^{\circ}$ angle, as illustrated in Fig.~\ref{fig:TopviewH}. Thus, the actual physical positions of $\mathrm{AP}_{i,j}$ and $U_0$ are given by $\mathbf{L}_{i,j} = \left(\left(i+\frac{j}{2}\right)d_{\mathrm{AP}},\frac{\sqrt{3}j}{2}d_{\mathrm{AP}}\right)$ and $\mathbf{L}_{U_0} = \left(\left(x_0+\frac{y_0}{2}\right)d_{\mathrm{AP}},\frac{\sqrt{3}y_0}{2}d_{\mathrm{AP}}\right)$, respectively, where $0\leq y_0\leq x_0\leq\frac{1}{2}$ and $x_0+\frac{y_0}{2}\leq\frac{1}{2}$.
\end{itemize}
Based on the aforementioned coordinate representation, the horizontal distance between $U_0$ and $\mathrm{AP}_{i,j}$ is given by 
\begin{align}
    d_{i,j} = ||\mathbf{L}_{U_0}-\mathbf{L}_{i,j}||.
\end{align}
For notational convenience, we introduce two constants $c_1$ and $c_2$, defined as
\begin{align}
    c_1 =\left\{
    \begin{aligned}
        &0,&&\textrm{ for square grid,}\\
        &\frac{1}{2},&&\textrm{ for hexagonal grid,}
    \end{aligned}\right.
\end{align}
and
\begin{align}
    c_2 =\left\{
    \begin{aligned}
        &1,&&\textrm{ for square grid,}\\
        &\frac{\sqrt{3}}{2},&&\textrm{ for hexagonal grid,}
    \end{aligned}\right.
\end{align}
respectively. Based on them, the physical positions of $\mathrm{AP}_{i,j}$ and $U_0$ can be unified as
\begin{align}
    \mathbf{L}_{i,j} = \bigl( (i + c_1 j) d_{\mathrm{AP}},\, c_2 j d_{\mathrm{AP}} \bigr),
\end{align}
and
\begin{align}
    \mathbf{L}_{U_0} = \bigl( (x_0 + c_1 y_0) d_{\mathrm{AP}},\, c_2 y_0 d_{\mathrm{AP}} \bigr),
\end{align}
respectively. Accordingly, the horizontal distance between $U_0$ and $\mathrm{AP}_{i,j}$ is expressed as
\begin{align}
    d_{i,j} = \sqrt{ \bigl( x_0 - i + c_1 (y_0 - j) \bigr)^2 + c_2^2 (y_0 - j)^2 } .
\end{align}
    

\begin{figure}[t]
    \centering
        \subfigure[The top view.]{
            \includegraphics[width=0.4\columnwidth]{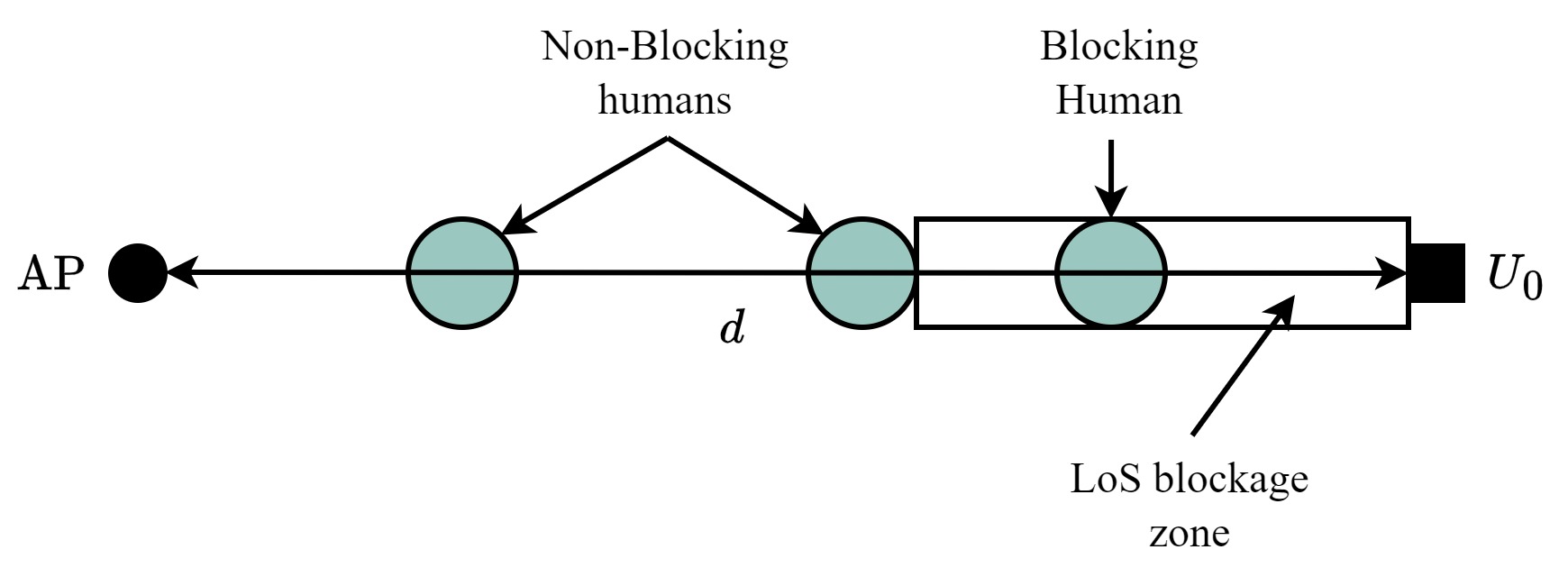}
            }\label{fig:TopviewHB}
        \subfigure[The vertical view.]{
            \includegraphics[width=0.4\columnwidth]{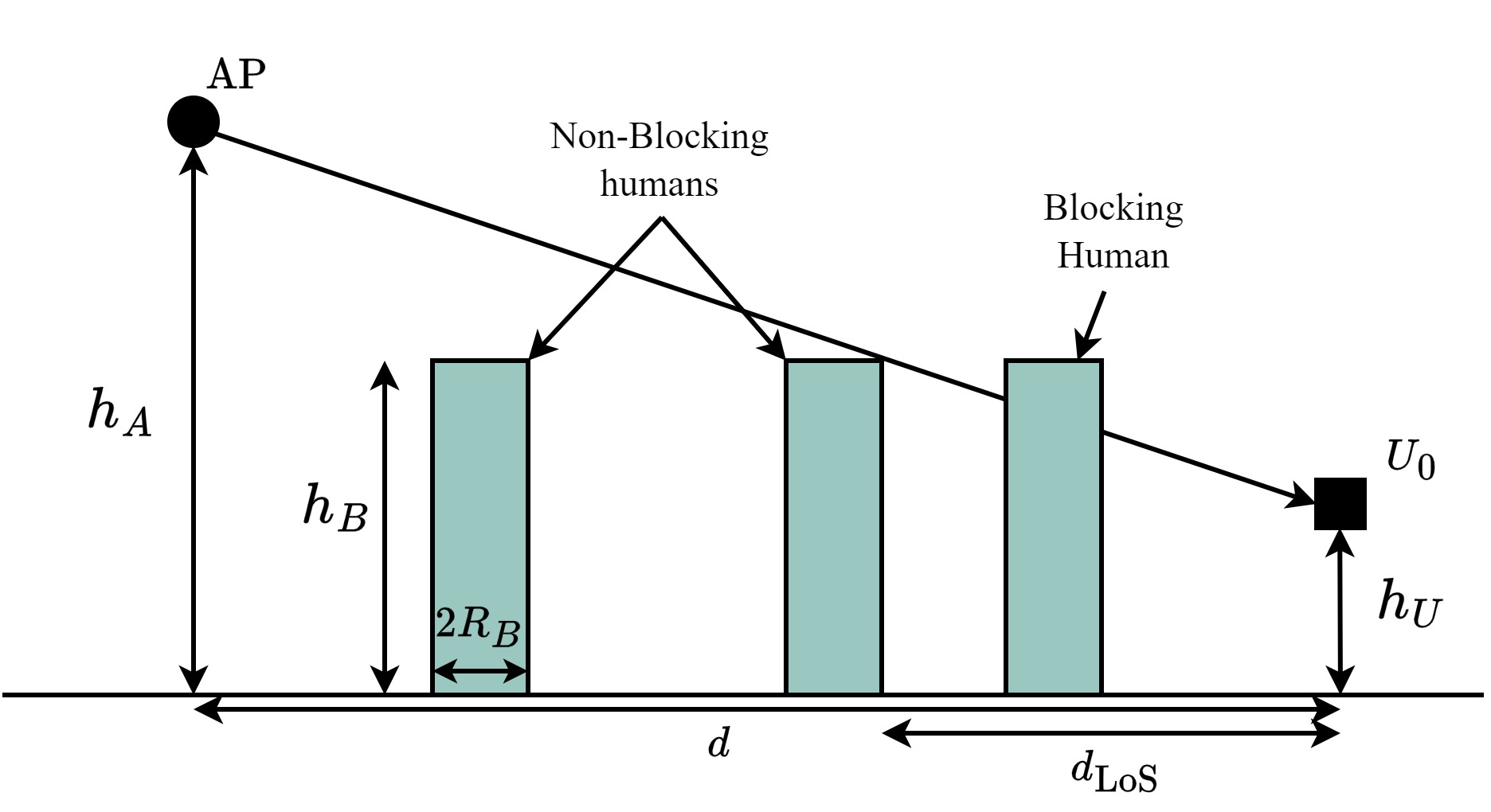}
            }\label{fig:VertviewHB}
            \vspace{-0.5em}
        \caption{Top and vertical views of human blockage for an AP-UE link.}
        \label{fig:HB}\vspace{-1em}
\end{figure}

With the AP–UE geometry and relevant distances, we next describe the blockage mechanisms in the considered system model. We first model the blockage caused by human bodies. Following state-of-the-art studies \cite{Wu2021TWC,Tang2025Tcom}, each human body is modeled as a vertical cylinder with radius $R_B$ and height $h_B$, whose bottom center follows a two dimensional Poisson point process with density $\lambda_B$. As illustrated in Fig.~\ref{fig:HB}, if the bottom center of a human body lies within the LoS blockage zone, the corresponding AP–UE transmission link is blocked. Conversely, if a human body is entirely outside the blockage zone, the AP–UE link is considered to be LoS, and the transmitted signal reaches the UE. We denote $\mathcal{B}_{i,j}^H$ as the human blockage indicator, where $\mathcal{B}_{i,j}^H=1$ if $\mathrm{AP}_{i,j}$ is blocked by the human bodies and $0$ otherwise. According to \cite{Wu2021TWC}, the probability that the link between $\mathrm{AP}_{i,j}$ and the typical UE is unblocked by a human body is given by $p_{i,j}^{\mathrm{H}} = \mathrm{Pr}(\mathcal{B}_{i,j}^H=0)=\exp(-\alpha d_{i,j})$, where $\alpha \delequal 2\lambda_B R_B (h_B-h_U)/\Delta h$ and $\Delta h \delequal h_A-h_U$.

 In addition to human blockages, wall blockages are also incorporated into the considered system model. Particularly, we employ the tractable Manhattan Poisson line processes (MPLPs) to describe the wall blockage model in the indoor environment, as in \cite{Wu2021TWC}, which accurately captures the grid like structure of indoor walls while maintaining flexibility for different room configurations. It is emphasized that the MPLP-based wall model serves as a statistical abstraction to evaluate performance averaged over a population of possible indoor layouts, rather than to represent a specific deterministic building configuration. Specifically, walls are oriented at either $0$ or $\pi/2$ angles to ensure that they are parallel or perpendicular to each other. The walls are modeled as two independent MPLPs, whose centers are distributed along the x-axis and the orthogonal direction of the x-axis on the ceiling plane, following independent one-dimensional PPPs, with density $\lambda_W$. We denote $\mathcal{B}_{i,j}^W$ as the wall blockage indicator, where $\mathcal{B}_{i,j}^W=1$ if the link from $\mathrm{AP}_{i,j}$ to $U_0$ is blocked by walls and $0$ otherwise. The expected number of vertical walls intersecting the link from $\mathrm{AP}_{i,j}$ to $U_0$ is proportional to the horizontal distance $d_{i,j,x} = d_{\mathrm{AP}}|x_0-i+c_1(y_0-j)|$, while the expected number of horizontal walls intersecting the link is proportional to $d_{i,j,y} = c_2 d_{\mathrm{AP}}| y_0-j|$. Accordingly, the probability that the LoS link from $\mathrm{AP}_{i,j}$ to $U_0$ is not blocked by walls is given by $p_{i,j}^{\mathrm{W}} =\mathrm{Pr}(\mathcal{B}_{i,j}^W=0)
= \exp\left(-\lambda_W (d_{i,j,x}+d_{i,j,y})\right)$. 

To jointly account for human and wall blockages, we define $\mathcal{B}_{i,j}$ as the overall blockage indicator of $\mathrm{AP}_{i,j}$, where $\mathcal{B}_{i,j}=1$ if the link is blocked by either human bodies or walls, and $\mathcal{B}_{i,j}=0$ otherwise. Following \cite{Shafie2021JSAC}, the probability that the link between $\mathrm{AP}_{i,j}$ and $U_0$ is not blocked is given by 
\begin{align}
p_{i,j} = \mathrm{Pr}(\mathcal{B}_{i,j}=0)= p_{i,j}^{H}p_{i,j}^{W} = \exp(-\chi_{i,j}),
\end{align}
where $\chi_{i,j} = \lambda_W (d_{i,j,x}+d_{i,j,y}) + \alpha d_{i,j}$.

\subsection{Antenna Model and Beam Training}

In this work, we assume that both APs and UEs are equipped with planar antennas to enhance the received signal power and compensate for the severe path loss inherent in THz propagation. Specifically, each AP is equipped with an $N_A\times N_A$ planar antenna array, while each UE is equipped with an $N_U\times N_U$ planar antenna array, both with an element spacing of $z=\frac{\lambda}{2}$, enabling narrow beamwidths and high antenna gains.

To establish reliable AP–UE links, each AP performs 3D beam training using a set of pyramid shaped beams with azimuth and elevation beamwidths denoted by $\omega_{H,T}$ and $\omega_{V,T}$, respectively \cite{tang2026impactpointingerrorcoverage}. Owing to the symmetric planar antenna structure at the AP, symmetric beamwidths are assumed in the two angular domains, such that $\omega_{H,T}=\omega_{V,T}=\omega_{T}$. During beam training, the entire azimuth and elevation angular ranges are partitioned into discrete sectors, and a hierarchical beam training procedure is adopted \cite{Ning2022twc}. At each training stage, the AP transmits a predefined set of training beams to identify the beam that maximizes the received signal power at the UE. To reduce training overhead, a coarse-to-fine beam training strategy is employed, in which $N_{\mathrm{ct}}$ candidate beams are transmitted concurrently at each training stage, followed by progressive angular refinement across stages. Since the UE may be located anywhere within the selected training beam sector, we assume that the UE location is uniformly distributed within the coverage region of each training beam. Consequently, the azimuth and elevation angular offsets between the AP beam steering direction and the actual UE direction can be modeled as independent and identically distributed random variables, uniformly distributed over $(-\omega_T,\omega_T)$.

\begin{figure}[t]
    \centering
        \subfigure[The top view.]{
            \includegraphics[width=0.4\columnwidth]{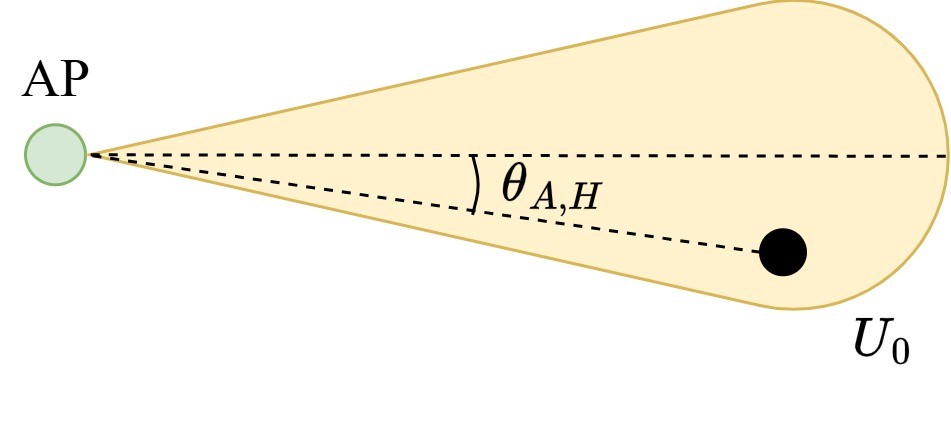}
            }\label{fig:TopviewB}
        \subfigure[The vertical view.]{
            \includegraphics[width=0.4\columnwidth]{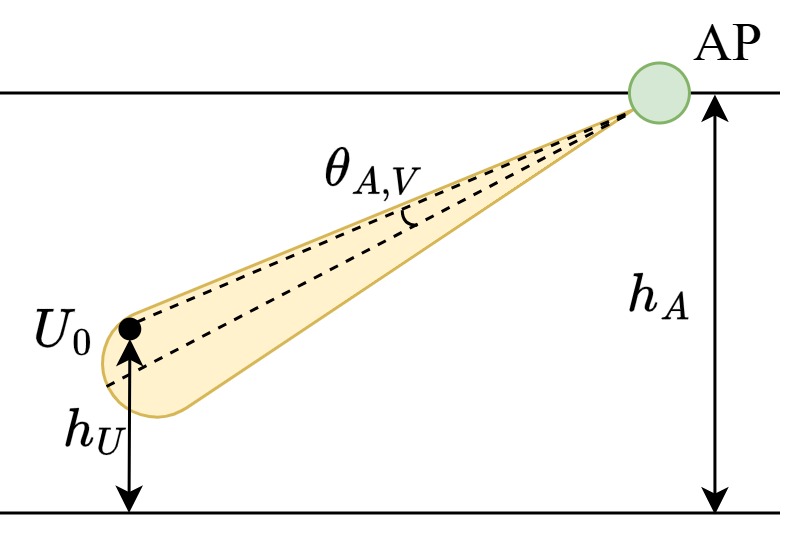}
            }\label{fig:VertviewB}\vspace{-0.5em}
        \caption{Top and vertical views of the pointing error for an AP-UE link.}\vspace{-1em}
        \label{fig:Beam}
\end{figure}

Owing to the random location of the UE within the selected beam sector, residual beam misalignment may occur after beam training, leading to a reduction in the effective received signal power. This residual pointing error is explicitly incorporated into the system analysis. According to \cite{Dabiri2022wcl}, the antenna gain in the direction specified by the elevation angle $\theta_q$ and azimuth angle $\phi_q$ can be expressed as $G_{q}(\theta_q,\phi_q)=G_{q,\max}H_{pe,q}(\theta_q,\phi_q)$, where $G_{q,\max}\approx \pi N_A^2$ represents the maximum antenna gain and angular deviation $\theta_q\!=\!\arctan\left(\!\sqrt{\tan^2(\theta_{q,V})\!+\!\tan^2(\theta_{q,H})}\!\right)$, with $\theta_{q,V}$ and $\theta_{q,H}$ representing the angles from the target direction in the horizontal and vertical planes, respectively, as shown in Fig.~\ref{fig:Beam}. Here, $q\in\{A_{i,j},U_0\}$, where $A_{i,j}$ corresponds to $\mathrm{AP}_{i,j}$ and $U_0$ for the $\mathrm{AP}_{i,j}$-$U_0$ link. The pointing error loss $H_{pe,q}(\theta_q,\phi_q)$ is given by
\begin{align}
    &H_{pe,q}(\theta_q,\phi_q) \notag\\
    &\!\!=\!\left(\!\frac{\sin\left(\frac{N_{q}\pi\sin(\theta_q)\sin(\phi_q)}{2}\right)\sin\left(\frac{N_{q}\!\pi\sin(\theta_q)\cos(\phi_q)}{2}\right)}{N_{q}\sin\left(\frac{\pi\sin(\theta_q)\!\sin(\phi_q)}{2}\right)\!N_{q}\!\sin\left(\frac{\pi\sin(\theta_q)\!\cos(\phi_q)}{2}\right)}\!\right)^2.
\end{align}
For small angular deviations, $\theta_q$ can be approximated by $\theta_q = \sqrt{\theta_{q,V}^2+\theta_{q,H}^2}$. Since APs are typically equipped with a substantially larger number of antenna elements than UEs, they can generate much narrower beams. As a result, the overall pointing error is dominated by the transmit beam misalignment at the AP, while the pointing error introduced by the UE antenna is neglected.

\subsection{Channel Model}

Signal propagation at THz frequencies is primarily governed by distance dependent large scale attenuation, including spreading loss and molecular absorption. By jointly accounting for these effects, the received power at $U_0$ from $AP_{i,j}$ is expressed as
\begin{align}
    P_{i,j} =  \left\{
    \begin{aligned} 
    &P_t G_{i,j} H_{i,j}, &&\textrm{if } \mathcal{B}_{i,j}=0,\\
    &0, &&\textrm{otherwise,}
    \end{aligned}\right.
\end{align}
where $P_t$ is the transmit power, $G_{i,j}=G_{{i,j},A}G_{{i,j},U}$ is the combined effective antenna gain, with $G_{{i,j},A}$ and $G_{{i,j},U}$ representing the antenna gains at $\mathrm{AP}_{i,j}$ and $U_0$ for the $\mathrm{AP}_{i,j}$-$U_0$ link, respectively, and $H_{i,j}$ denotes large-scale pathloss for the $\mathrm{AP}_{i,j}$-$U_{0}$ link. The large scale fading gain from $\mathrm{AP}_{i,j}$ to $U_0$, separated by a horizontal distance $d_{i,j}$ in a 3D indoor environment, is modeled as $H_{i,j} = \xi W(d_{i,j})$, where $\xi\delequal \frac{c^2}{(4\pi f)^2}$, $c=3\times10^8$ m/s is the light speed, $f$ is the operating frequency, $W(d_{i,j})=\frac{1}{d_{i,j}^2+(h_A-h_U)^2}\exp\left(-\epsilon(f)\sqrt{d_{i,j}^2+(h_A-h_U)^2}\right)$, and $\epsilon(f)$ is the molecular absorption coefficient of frequency $f$.

In the considered system, we adopt the nearest LoS AP association strategy, i.e., each UE associates with its nearest AP that has a LoS propagation path to the UE. We denote $\mathcal{A}_{i,j}$ as the event that the typical UE is associated with $\mathrm{AP}_{i,j}$ and define the set of interfering APs as $\Psi_{\mathrm{AP}_{i,j}}=\{\textrm{AP}_{m,n}|d_{m,n}\geq d_{i,j},(m,n)\neq(i,j)\}$. It is worth noting that NLoS paths are treated as fully blocked, which is a common simplification at indoor THz frequencies, e.g., 0.3 THz, to ensure analytical tractability despite potential reflected components.

To evaluate the system performance, we employ the coverage probability, denoted by $P_c$, as our performance metric. It is defined as the probability that the received signal-to-interference-plus-noise ratio (SINR) at $U_0$ exceeds a given threshold $\beta$, i.e., $P_c = \mathrm{Pr}(\mathrm{SINR}>\beta)$. Conditioned on the association event $\mathcal{A}_{i,j}$, the SINR of $U_0$ is given by
\begin{align}\label{eq:SINRequation}
    \mathrm{SINR} = \frac{P_{i,j}}{I_{i,j} + N_0}
    = \frac{P_t\xi G_{i,j}W(d_{i,j})}{\sum\limits_{\mathrm{AP}_{m,n}\in \Psi_{\mathrm{AP}_{i,j}}} P_{m,n} + N_0},
\end{align}
where $I_{i,j}$ denotes the aggregate interference from all non associated LoS APs and $N_0$ denotes the noise power.


\section{Performance Analysis}\label{Sec:Coverage}

In this section, we analyze the coverage probability of the typical UE $U_0$. Conditioned on the horizontal coordinate of $U_0$,
${\mathcal{O}}_{U_0} = (x_0, y_0)$, the coverage probability is expressed as
\begin{align}\label{eq:Pc}
    P_c 
    &= \mathbb{E}\!\left[ \Pr(\mathrm{SINR} > \beta \mid x_0, y_0) \right]\notag\\
    &= \sum_{d_{i,j}\leq R} 
    \Pr(\mathrm{SINR} > \beta \mid \mathcal{A}_{i,j})\,
    \Pr(\mathcal{A}_{i,j}),
\end{align}
where the conditioning on the horizontal location $(x_0,y_0)$ of $U_0$ is suppressed in the notation for brevity, and $\Pr(\mathcal{A}_{i,j} )$ denotes the probability that $U_0$ is associated with $\mathrm{AP}_{i,j}$. Conditioned on the association event $\mathcal{A}_{i,j}$, the conditional coverage probability is given by
\begin{align}\label{eq:PCovercal1}
    &\Pr(\mathrm{SINR}\! >\! \beta\! \mid\! \mathcal{A}_{i,j})
   \! =\! \Pr\!\left( \frac{H_{pe} \zeta_{i,j}}{I_{i,j}+N_0} >\! \beta 
    \bigg| \mathcal{A}_{i,j} \right),
\end{align}
where
\begin{align}
    \zeta_{i,j}
    =  P_t \xi G_{\max} W(d_{i,j})
\end{align}
and $G_{\max} = G_{A_{i,j},\max}G_{U_0,\max}$. To evaluate the coverage probability of $U_0$, we first characterize the effects of pointing error and wall blockages on the received signal and interference.

\vspace{-0.5em}
\subsection{Impact of Pointing Error and Wall Blockages}

Under correct finite-resolution beam training, the AP selects the transmit beam that covers the LoS sector toward the typical UE $U_0$. Due to the finite beamwidth of the training beams, residual beam misalignment remains even after beam training, which constitutes a dominant source of received signal power degradation. To enable tractable performance analysis, we characterize the statistical behavior of the pointing error loss $H_{pe}$ by deriving approximate expressions for its probability density function and cumulative distribution function, as summarized in the following lemma.

\begin{Lemma}\label{Lemma:1}
 The PDF and the CDF of pointing error loss, $f_{H_{pe}}(h_{pe})$ and $F_{H_{pe}}(h_{pe})$, are approximated as 
    \begin{align}\label{eq:fhpe}
        &f_{H_{pe}}(h) \approx \left\{
        \begin{aligned}
            &\frac{\pi \omega_A^2}{4\omega_{T}^2 h}, &&\textrm{if } \omega_1\leq h\leq 1,\\
            &\frac{\omega_A^2}{\omega_{T}^2 h}\left(\rho_1-\frac{\pi}{4}\right), &&\textrm{if } \omega_1^2\leq h< \omega_1,\\
            &0 , &&\textrm{otherwise,}
        \end{aligned}\right.
    \end{align}
    and
    \begin{align}\label{eq:Fchpe}
        &F_{H_{pe}}(h) \notag\\
        &\approx \left\{
        \begin{aligned}
            &1+\frac{\pi \omega_A^2 \ln h}{4\omega_{T}^2}, &&\textrm{if } {\omega_1\leq h\leq 1,}\\
            &1 + \frac{\omega_A^2 \ln h}{\omega_{T}^2 }\left(\rho_1 -\frac{\pi}{4}\right)-\rho_2, &&\textrm{if } \omega_1^2\leq h< \omega_1,\\
            &1 , &&\textrm{if }h> 1\\
            &0 , &&\textrm{otherwise,}
        \end{aligned}\right.
    \end{align}
    respectively, where $\rho_1=\arcsin\left(\frac{\omega_{T}}{\sqrt{-\omega_A^2\ln h}}\right)$, $\rho_2=\frac{\sqrt{-\omega_A^2 \ln h-\omega_{T}^2}}{\omega_{T}}$, $\omega_A = \frac{1.06}{N_A}$, and $\omega_1=e^{-\omega_{T}^2/\omega_A^2}$.
    \begin{IEEEproof}
        See Appendix~\ref{Appendix:A}.
    \end{IEEEproof}

\end{Lemma}

In addition to residual pointing error, wall blockages play a critical role in determining the propagation conditions of LoS AP-UE links in indoor environments. Under the MPLP wall model, blockage events across different AP-UE links are generally correlated. In particular, a single wall can simultaneously block multiple APs whose line-of-sight paths intersect the same wall segment. This correlation arises because APs located in similar directions relative to the UE may share common wall intersections. Consequently, the blockage states of different AP--UE links are not independent. In both square and hexagonal grid AP deployments, the joint unblocked probability of multiple APs depends on the extent of overlap between their horizontal and vertical projections toward the UE. Such spatial correlation has a direct impact on the statistics of aggregate interference and received signal strength, and must be explicitly accounted for in the coverage analysis.

While the marginal unblocked probability $p_{i,j}^{\mathrm{W}}$ characterizes the likelihood that the individual $\mathrm{AP}_{i,j}$-$U_0$ link is unblocked, it is insufficient to capture the joint behavior of multiple interfering links. To quantify the effect of correlated wall blockages, we analyze the covariance between the wall blockage indicators of different APs, which serves as a key ingredient for the subsequent interference analysis.

\begin{Lemma}\label{Lemma:Correlation}
    Considering two APs, $\mathrm{AP}_{i_1,j_1}$ and $\mathrm{AP}_{i_2,j_2}$, the covariance between their wall blockage indicators is given by
    \begin{align}\label{eq:covariance}
    \mathrm{Cov}(\mathcal{B}_{i_1,j_1},\mathcal{B}_{i_2,j_2}) = {p^{W}_{(i_1,j_1),(i_2,j_2)}-p^W_{i_1,j_1}p^W_{i_2,j_2}},
    \end{align}
    where $p^{W}_{(i_1,j_1),(i_2,j_2)} = \exp(-\lambda_W(U_{(i_1,j_1),(i_2,j_2),x}+U_{(i_1,j_1),(i_2,j_2),y}))$ denotes the joint probability that both links are unblocked by walls. The functions $U_{(i_1,j_1),(i_2,j_2),x}$ and $U_{(i_1,j_1),(i_2,j_2),y}$ are given by
    \begin{align}
        &U_{(i_1,j_1),(i_2,j_2),x} \notag\\
        &= \left\{
        \begin{aligned}
            &\max(d_{i_1,j_1,x}, d_{i_2,j_2,x}), && \textrm{if } \vec{d}_{i_1,j_1,x}\vec{d}_{i_2,j_2,x} > 0,\\
            &d_{i_1,j_1,x} + d_{i_2,j_2,x}, && \text{otherwise}
        \end{aligned}\right.
    \end{align}
    and 
    \begin{align}
        &U_{(i_1,j_1),(i_2,j_2),y} \notag\\
        &= \left\{
        \begin{aligned}
            &\max(d_{i_1,j_1,y}, d_{i_2,j_2,y}), && \textrm{if } \vec{d}_{i_1,j_1,y}\vec{d}_{i_2,j_2,y} > 0,\\
            &d_{i_1,j_1,y} + d_{i_2,j_2,y}, && \text{otherwise},
        \end{aligned}\right.
    \end{align}
    respectively, where $\vec{d}_{i,j,x} = x_0 - i + c_1(y_0 - j)$ and $\vec{d}_{i,j,y} = (y_0 - j)$.
    \begin{IEEEproof}
         The result follows from the geometric properties of the MPLP wall model. For two APs, the joint probability that both links are unblocked corresponds to having no vertical walls in the union of their horizontal projections, and no horizontal walls in the union of their vertical projections, which results in $p_{(i_1,j_1),(i_2,j_2)}^{\mathrm{W}}$. The covariance of the Bernoulli blockage indicators is given by
        \begin{align}\label{eq:CovCalcu1}
        &\mathrm{Cov}(\mathcal{B}_{i_1,j_1}, \mathcal{B}_{i_2,j_2}) = \notag\\
        &{\Pr(\mathcal{B}_{i_1,j_1}=1, \mathcal{B}_{i_2,j_2}=1) - \Pr(\mathcal{B}_{i_1,j_1}=1) \Pr(\mathcal{B}_{i_2,j_2}=1)}.
        \end{align}
        Substituting $\Pr(\mathcal{B}_{i,j}=1) = 1 - p_{i,j}^{\mathrm{W}}$ and $\Pr(\mathcal{B}_{i_1,j_1}=1, \mathcal{B}_{i_2,j_2}=1) = 1 - p_{i_1,j_1}^{\mathrm{W}} - p_{i_2,j_2}^{\mathrm{W}} + p_{(i_1,j_1),(i_2,j_2)}^{\mathrm{W}}$ into \eqref{eq:CovCalcu1} yields the results in \eqref{eq:covariance}.
    \end{IEEEproof}
\end{Lemma}

\textit{Remark 1:} Lemma~\ref{Lemma:Correlation} shows that wall blockages introduce spatial correlation among AP--UE links, governed by the geometric overlap of their propagation paths. The APs located in similar directions relative to the UE tend to share common wall blockages, resulting in positively correlated blockage states, whereas links with disjoint projections exhibit weaker correlation. This correlation cannot be captured by marginal unblocked probabilities alone and plays a critical role in determining interference fluctuations. Accounting for such correlation is therefore essential for accurate evaluation of aggregate interference and reliable coverage analysis in grid structured indoor THz systems.

\subsection{Coverage Analysis}

To evaluate the coverage probability, we first characterize the antenna gains of interfering signals, which form the basis for the subsequent interference analysis. To simplify interference modeling, we adopt the cone antenna model proposed in \cite{Shafie2021JSAC}. Under this model, the antenna gain of an interfering signal received at $U_0$ from a LoS non-associated AP, $\mathrm{AP}_{m,n}$, is approximated by the side-lobe gain, obtained as $G_{m,n} =G_{\mathrm{S}}=  G_{A_{m,n},\mathrm{S}}G_{U_0,\mathrm{S}}$, with
\begin{align}
    G_{q,\mathrm{S}} = \frac{\pi-N_q^2\pi\arcsin\left(\tan\left(\frac{\phi_{q,V}}{2}\right)\tan\left(\frac{\phi_{q,H}}{2}\right)\right)}{\pi-\arcsin\left(\tan\left(\frac{\phi_{q,V}}{2}\right)\tan\left(\frac{\phi_{q,H}}{2}\right)\right)},
\end{align}
where $\phi_{q,V}$ and $\phi_{q,H}$ are the vertical and horizontal main lobe beamwidths, respectively \cite{tang2026impactpointingerrorcoverage}. We next investigate the behavior of the aggregate interference generated by non-associated APs. In particular, we examine whether the contributions from distant APs remain bounded, which is a key prerequisite for developing a tractable interference model and coverage analysis.

\begin{Lemma}\label{Lemma:3}
    When the typical UE, $U_0$, is associated with $\mathrm{AP}_{i,j}$, the aggregate interference $I_{i,j}$ is convergent. Moreover, its mean and variance, denoted by $\mu_{I_{i,j}}$ and $\sigma_{I_{i,j}}^2$, are given by
    \begin{align}\label{eq:averageI}
            \mu_{I_{i,j}} = P_t G_{\mathrm{S}}\xi&\sum\limits_{(m,n)\in \Psi_{i,j}  }p_{m,n}^H  \notag\\
            &\times\mathbb{E}_{\mathcal{B}^W}\left[\overline{\mathcal{B}}_{m,n}^W\big|\overline{\mathcal{B}}_{i,j}^W\!=\!1\right] W(d_{m,n}),
        \end{align}
       and  
        \begin{align}\label{eq:varianceI}
            &\sigma_{I_{i,j}}^2 = \sum_{(m,n)\in \Psi_{i,j}} p_{m,n}^H \mathbb{E}_{\mathcal{B}^W}\left[\overline{\mathcal{B}}_{m,n}^W\big|\overline{\mathcal{B}}_{i,j}^W=1\right] \notag\\
            &\times\left(1 - p_{m,n}^H \mathbb{E}_{\mathcal{B}^W}\left[\overline{\mathcal{B}}_{m,n}^W\big|\overline{\mathcal{B}}_{i,j}^W=1\right]\right) (P_t G_{\mathrm{S}}\xi W(d_{m,n}))^2 \notag\\
        &+ \sum_{\substack{(m_1,n_1)\neq (m_2,n_2)}} \mathrm{Cov}\left(\overline{\mathcal{B}}_{m_1,n_1}^W, \overline{\mathcal{B}}_{m_2,n_2}^W\big|\overline{\mathcal{B}}_{i,j}^W=1\right) \notag\\
        &\times p_{m_1,n_1}^H p_{m_2,n_2}^H (P_t G_{\mathrm{S}}\xi)^2 W(d_{m_1,n_1}) W(d_{m_2,n_2}),
        \end{align}
        respecively, where $\mathrm{Cov}(B_{m_1,n_1}^W, B_{m_2,n_2}^W)$ is given in \eqref{eq:covariance} in Lemma~\ref{Lemma:Correlation}.
\begin{IEEEproof}
 See Appendix.~\ref{Appendix:B}.
\end{IEEEproof}
    
\end{Lemma}

Lemma~\ref{Lemma:3} shows that the aggregate interference remains finite despite the infinite AP lattice. This is due to the combined effects of distance dependent path loss, molecular absorption, and blockage, which introduce an effective exponential decay and render the contribution of distant APs negligible. As a result, the total interference can be accurately approximated by restricting attention to a finite neighborhood around the typical UE, thereby substantially simplifying both analytical derivations and numerical evaluations. Moreover, the existence of finite first and second order moments enables the interference to be characterized by its mean in \eqref{eq:averageI} and variance in \eqref{eq:varianceI}, leading to a tractable and reliable coverage probability analysis in grid structured THz systems.

It is worth noting that both the desired signal power and the aggregate interference depend critically on the AP associated with the typical UE. Therefore, characterizing the association probability is a fundamental step toward evaluating the overall coverage probability, which is addressed in the following Lemma.

\begin{Lemma}\label{lem:PA_exact}
The probability that the typical UE $U_0$ is associated with $\mathrm{AP}_{i,j}$, denoted by $\Pr(\mathcal{A}_{i,j})$, is given by
\begin{align}\label{eq:PrAij}
\Pr(\mathcal{A}_{i,j}) 
=&\; p^H_{i,j} \sum_{\Phi \subseteq \overline{\Psi}_{i,j}} (-1)^{|\Phi|} 
\left( \prod_{(m,n) \in \Phi} p^H_{m,n} \right) \notag\\
&\times \exp\!\big(-\lambda_W(U_{(i,j),\Phi,x}+U_{(i,j),\Phi,y})\big),
\end{align}
where $\overline{\Psi}_{i,j}$ denotes the set of APs that are closer to the typical UE than $\mathrm{AP}_{i,j}$, and $\Phi$ is an arbitrary subset of $\overline{\Psi}_{i,j}$ arising from the inclusion–exclusion expansion. The terms $U_{(i,j),\Phi,x}$ and $U_{(i,j),\Phi,y}$ represent the union lengths of the projections of all links from $U_0$ to the APs in the set $\{(i,j)\}\cup\Phi$ onto the $x$- and $y$-axes, respectively, given by
\begin{align}
U_{(i,j),\Phi,x} &= \max_{(m_1,n_1),(m_2,n_2)\in\{(i,j)\}\cup\Phi} U_{(m_1,n_1),(m_2,n_2),x} ,
\end{align}
and
\begin{align}
U_{(i,j),\Phi,y} &= \max_{(m_1,n_1),(m_2,n_2)\in\{(i,j)\}\cup\Phi} U_{(m_1,n_1),(m_2,n_2),y},
\end{align}
respectively. 
\begin{IEEEproof}
See Appendix~\ref{Appendix:Lemma4}.
\end{IEEEproof}

\end{Lemma}

Based on Lemma~\ref{lem:PA_exact}, the association probability, i.e., the probability that at least one AP provides a LoS link and covers the typical UE, is obtained by summing $\Pr(\mathcal{A}_{i,j})$ over all APs within the coverage radius, given by
\begin{align}
    P_{\mathrm{Assoc}} = \sum_{(i,j):\, d_{i,j} \leq R} \Pr(\mathcal{A}_{i,j}).
\end{align}

The exact expression in Lemma~\ref{lem:PA_exact} captures the impact of correlated wall blockages through the joint wall-unblocked probability $p^W_{(i,j)\cup\Phi}$. Even when the AP–UE link distances are relatively short or the wall density $\lambda_W$ is moderate, a single wall may simultaneously obstruct multiple AP–UE links, inducing non-negligible correlation among blockage events. As a result, the commonly adopted independence assumption for wall blockages may lead to inaccurate association and coverage characterizations \cite{Wu2021TWC,Shafie2021JSAC,Kouzayha2023twc}.


By combining the association probability in Lemma~\ref{lem:PA_exact}, the interference statistics in Lemma~\ref{Lemma:3}, and the statistical characterization of the pointing error in Lemma~\ref{Lemma:1}, we derive the coverage probability of the typical UE.

\begin{Theorem}\label{Theorem:Coverage}
    The coverage probability of the typical UE $U_0$ can be expressed as
    \begin{align}\label{eq:coverage}
        P_c \approx&\sum\limits_{d_{i,j}\leq R}\mathrm{Pr}(\mathcal{A}_{i,j}) \Bigg(1 -  F_{H_{pe}}\left(\frac{(\mu_{I_{i,j}}+N_0)\beta}{\zeta_{i,j}}\right) \notag\\
            &- \frac{ \beta^2\sigma_{I_{i,j}}^2}{2\zeta_{i,j}^2}  f_{H_{pe}}'\left(\frac{(\mu_{I_{i,j}}+N_0)\beta}{\zeta_{i,j}}\right)\Bigg),
    \end{align}
    where $\mu_{I_{i,j}}$ and $\sigma^2_{I_{i,j}}$ are defined in Lemma~\ref{Lemma:3}, and $f_{H_{pe}}'(\cdot)$ denotes the first-order derivative of $f_{H_{pe}}(\cdot)$, given by
        \begin{align}\label{eq:ffirstod}
        &f_{H_{pe}}'(h) \approx \notag\\
        &\left\{
        \begin{aligned}
            &-\!\frac{\pi \omega_A^2}{4\omega_{T}^2 h^2}, &&\textrm{if } \omega_1\!\leq\! h\!\leq\! 1,\\
            &-\!\frac{\pi \omega_A^2}{\omega_{T}^2 h^2}\left(\rho\!-\!\frac{\pi}{4}\right)\! -\! \frac{\omega_A^2}{2\omega_{T}^2h^2\rho_2\ln h}, &&\textrm{if } \omega_1^2\!\leq\! h\!<\! \omega_1,\\
            &0 , &&\textrm{otherwise.}
        \end{aligned}\right.
    \end{align}
    \begin{IEEEproof}
    See Appendix~\ref{Appendix:C}.
    \end{IEEEproof}
\end{Theorem}

Based on the above analysis, the average coverage probability over all possible UE locations can be obtained by averaging the conditional coverage probability over the fundamental region, given by
\begin{align}
    \overline{P}_c \!=\! \int_{0}^{\frac{1}{2}}\int_{0}^{y_{\mathrm{th}}} \sum\limits_{d_{i,j}\leq R}\underbrace{\mathrm{Pr}\left(\mathrm{SINR}>\beta|\mathcal{A}_{i,j}\right)\mathrm{Pr}(\mathcal{A}_{i,j})}_{P_c}\mathrm{d}y\mathrm{d}x,
\end{align}
where $y_{\mathrm{th}} = \min\{x,1-4c_1 x\}$. However, this expression is difficult to further simplify into a closed-form result. This is because the association probability $\Pr(\mathcal{A}_{i,j})$ depends strongly on the UE location. In addition, the inclusion–exclusion expansion in \eqref{eq:PrAij} introduces combinatorial complexity that grows exponentially with the number of APs, while the exponential terms depend on the wall blockage configuration of each subset, leading to strong spatial correlations between the APs and the UE. These factors together render a simplified closed-form expression for the coverage probability analytically intractable.


\subsection{Beam Training Overhead}

During the beam training phase, increasing the number of concurrent training beams $N_{\mathrm{ct}}$ reduces the total number of required training stages, but at the cost of increased intra-AP interference among simultaneously transmitted beams. From a hardware perspective, the maximum degree of parallel beam training is constrained by the number of RF chains available at the AP, since each RF chain can support at most one independent training beam. However, this hardware limitation alone does not fully determine the feasible value of $N_{\mathrm{ct}}$.

In practice, excessive parallelization of training beams leads to severe intra-AP interference, which degrades the reliability of beam detection. As a result, there exists a fundamental trade-off between training efficiency and detection reliability. Specifically, $N_{\mathrm{ct}}$ must be selected such that the received SINR associated with each training beam remains above a prescribed threshold. We characterize this requirement by a fixed SINR threshold $\beta_{\mathrm{ct}}$, which represents the minimum signal quality required for correct beam detection under a given training and receiver processing scheme.

To ensure reliable beam training throughout the coverage area of radius $R$, we consider a worst-case scenario in which the received SINR at the cell boundary must satisfy the channel-training requirement. Moreover, at the boundary of a training beam, the effective pointing error loss is given by $H_{pe}=\omega_1^2$. Accordingly, we impose the following SINR constraint
\begin{align}\label{eq:SINRrequirement}
    \mathrm{SINR}(R) 
    = \frac{\frac{P_t}{N_{\mathrm{ct}}} \omega_1^2 G_{\mathrm{max}}\xi W(R)}
           {I_{\mathrm{intra}} + I_{\mathrm{inter}}(R)  + N_0}
    \ge \beta_{\mathrm{ct}},
\end{align}
where $I_{\mathrm{intra}}$ denotes the intra-AP interference caused by the simultaneous transmission of the remaining training beams from the same AP, given by
\begin{align}\label{eq:Iintra}
    I_{\mathrm{intra}} = \frac{(N_{\mathrm{ct}}-1)P_t}{N_{\mathrm{ct}}}G_{\mathrm{S}}\xi W(R),
\end{align}
and $I_{\mathrm{inter}}(R)$ represents the aggregate interference from all other APs located at distances larger than $R$, expressed as
\begin{align}
    I_{\mathrm{inter}}(R) 
    = \sum_{(m,n): d_{m,n}>R} 
      P_t  \xi p_{m,n}^H W(d_{m,n}).
\end{align}
By enforcing the SINR requirement in \eqref{eq:SINRrequirement} and accounting for both intra-AP and inter-AP interference, an upper bound on the number of concurrent training beams can be obtained, as summarized in the following Lemma.

\begin{Lemma}\label{Lemma:5}
The maximum number of concurrent training beams is upper-bounded by
\begin{align}\label{eq:Nctmax}
N_{\mathrm{ct}}^{\max} = \left\lfloor\frac{\omega_1^2G_{\textrm{max}}+\beta_{\mathrm{ct}}G_{\mathrm{S}}}{\beta_{\mathrm{ct}}\left(G_{\mathrm{S}}+\eta\right)}\right\rfloor,
\end{align}
where $\eta = \frac{I_{\mathrm{inter}}(R)+N_0}{P_t \xi W(R)}$. The aggregate inter-AP interference can be approximated as
    \begin{align}\label{eq:Iapprox}
    I_{\mathrm{inter}}(R) \approx &\frac{2\pi P_{t} G_{\mathrm{S}} \xi R e^{- \alpha  R} W(R)}{d_{AP}^2(\alpha + \epsilon(f) )} I_{0}\left(\sqrt{2}\lambda_W R\right),
    \end{align}
where $I_{0}(\cdot)$ denotes the modified Bessel function of the first kind with zero order.
\begin{IEEEproof}
    See Appendix~\ref{Appendix:Lemma5}.
\end{IEEEproof}
\end{Lemma}

Based on Lemma~\ref{Lemma:5}, the total number of beam training stages required to complete the hierarchical beam training procedure is given by
\begin{align}\label{eq:NBT}
    N_{\mathrm{BT}}
    = \left\lceil
    \log_{N_{\mathrm{ct}}}
    \frac{4\pi \arctan\!\left(\frac{R}{\Delta h}\right)}
    {\omega_T^2}
    \right\rceil,
\end{align}
where $N_{\mathrm{ct}}=\min\{N_{\mathrm{ct}}^{\max},N_{\mathrm{RF}}\}$ and $N_{\mathrm{RF}}$ denotes the number of RF chains available at the AP.

\textit{Remark 2:} Lemma~\ref{Lemma:5} reveals how the maximum number of concurrent training beams $N_{\mathrm{ct}}^{\max}$ depends on key system parameters. In particular, a larger mainlobe gain $G_{\mathrm{max}}$ or improved pointing accuracy allows a higher degree of parallel beam training. Nevertheless, the effective number of concurrently trainable beams is fundamentally constrained by the number of RF chains at the AP. Moreover, achieving higher antenna gain and improved pointing accuracy typically requires a larger antenna array, which results in narrower beams and hence a smaller beamwidth $\omega_T$. As indicated by \eqref{eq:NBT}, a reduced beamwidth increases the total number of candidate beams required to cover the service area, thereby increasing the number of beam training stages. At the same time, a larger antenna array enhances training parallelism by permitting a larger $N_{\mathrm{ct}}$ within the RF-chain constraint. These competing effects imply that the total beam training overhead is a non-monotonic function of the antenna array size. This observation highlights the importance of jointly designing the antenna array configuration and the beam training strategy to minimize overall training latency in indoor THz systems.

\section{Numerical Results}\label{Sec:Num}

\begin{table*} 
\centering
\caption{Value of System Parameters Used in Section~\ref{Sec:Num}.}\vspace{-0.5em}
\begin{tabular}{|l|l|l|l|}
\hline
    \textbf{Parameter Type} &\textbf{Parameter} & \textbf{Symbol} & \textbf{Value} \\
    \hline
    \multirow{3}*{AP and UE}&Height of APs, UEs  &  $h_A$, $h_U$ & $3$ m, $1.3$ m\\\cline{2-4}
    \multirow{3}*{}&Inter-AP distance and coverage boundary  &  $d_{\mathrm{AP}}$, $R_A$ & $15$ m, $15$ m\\\cline{2-4}
    \multirow{3}*{}&Number of antennas  &  $N_A$, $N_U$ & $16$, $2$\\\hline
     \multirow{2}*{Wall and human blockages}&Density of wall and human blockages  &  $\lambda_W$, $\lambda_B$ & $0.02$ $\mathrm{m}^{-1}$, $0.1$ $\mathrm{m}^{-2}$ \\\cline{2-4}
    \multirow{2}*{}&Radius and height of human body  &  $R_B$, $h_B$& $0.25$ $\mathrm{m}$, $1.7$ m \\\hline
    \multirow{3}*{THz transmission} &Operating frequency and  bandwidth &  $f$, $B$ & $300$ GHz, $5$ GHz \\\cline{2-4}
     \multirow{3}*{} &Absorption coefficient &  $\epsilon(f)$ & $0.00143$ $\mathrm{m}^{-1}$ \\\cline{2-4}
     \multirow{3}*{}&Transmit power and AWGN power & $P_t$, $N_0$ & $5$ dBm, $-77$ dBm \\\hline
    \multirow{2}*{Beam training parameter}&Beamwidth and number of RF chains & $\omega_{T}$, $N_{\mathrm{RF}}$ & $0.0554$ rad, 6 \\\cline{2-4}
    \multirow{2}*{}&SINR requirement & $\beta_{\mathrm{ct}}$ & $10$ dB \\\hline
\end{tabular}
\label{tab:System_Para}\vspace{-1.5em}
\end{table*}

\begin{figure}[t]
    \centering
    \subfigure[The locations in square grid.]
        {    \label{fig:LocationS}        
            \includegraphics[width=0.4\columnwidth]{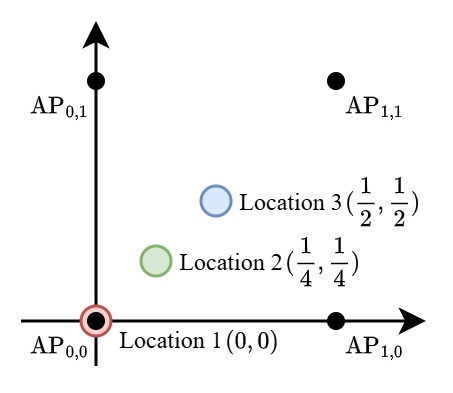}
            }
        \subfigure[The locations in hexagonal grid.]{\label{fig:LocationH}
             \includegraphics[width=0.43\columnwidth]{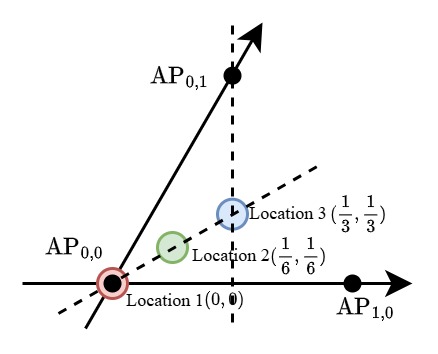}
            }\vspace{-0.5em}
        \caption{Illustration of the considered three locations of the typical UE, where the red point is \emph{Location~1} directly under $\mathrm{AP}_{0,0}$, the blue point is \emph{Location~3} at the farthest position from $\mathrm{AP}_{0,0}$, and the green point is \emph{Location~2} at the midpoint between red and blue points.}\vspace{-1.5em}
        \label{fig:Location}
\end{figure}

In this section, we first present numerical results to validate the analytical framework developed in Section~\ref{Sec:Coverage}. We then investigate the impact of key system parameters on the association probability, coverage probability, and the number of beam training stages, including the wall blockage density, grid structure, UE location, and beamwidth. To examine the effect of the UE location, we consider three representative scenarios, referred to as \emph{Location~1}, \emph{Location~2}, and \emph{Location~3}. Specifically, the typical UE is located (i) directly beneath $\mathrm{AP}_{0,0}$, with ${\mathcal{O}_{U_0}}=(0,0)$ for both square and hexagonal grid AP deployments, (ii) at the farthest location from $\mathrm{AP}_{0,0}$, given by ${\mathcal{O}_{U_0}}=(1/2,1/2)$ for the square grid and ${\mathcal{O}_{U_0}}=(1/3,1/3)$ for the hexagonal grid, and (iii) at the midpoint between the previous two locations, i.e., ${\mathcal{O}_{U_0}}=(1/4,1/4)$ for the square grid and ${\mathcal{O}_{U_0}}=(1/6,1/6)$ for the hexagonal grid, respectively, as illustrated in Fig.~\ref{fig:Location}. Simulation results are obtained using MATLAB by averaging over $10^8$ independent realizations, while the analytical results are evaluated numerically based on the derived expressions. Unless otherwise specified, the system parameters used in this section are summarized in Table~\ref{tab:System_Para}. These parameter values are consistent with those commonly adopted in the THz communication literature \cite{Shafie2021JSAC,Tang2025Tcom,Wu2021TWC}.

\begin{figure}[t]
    \centering
    \includegraphics[width=0.9\columnwidth]{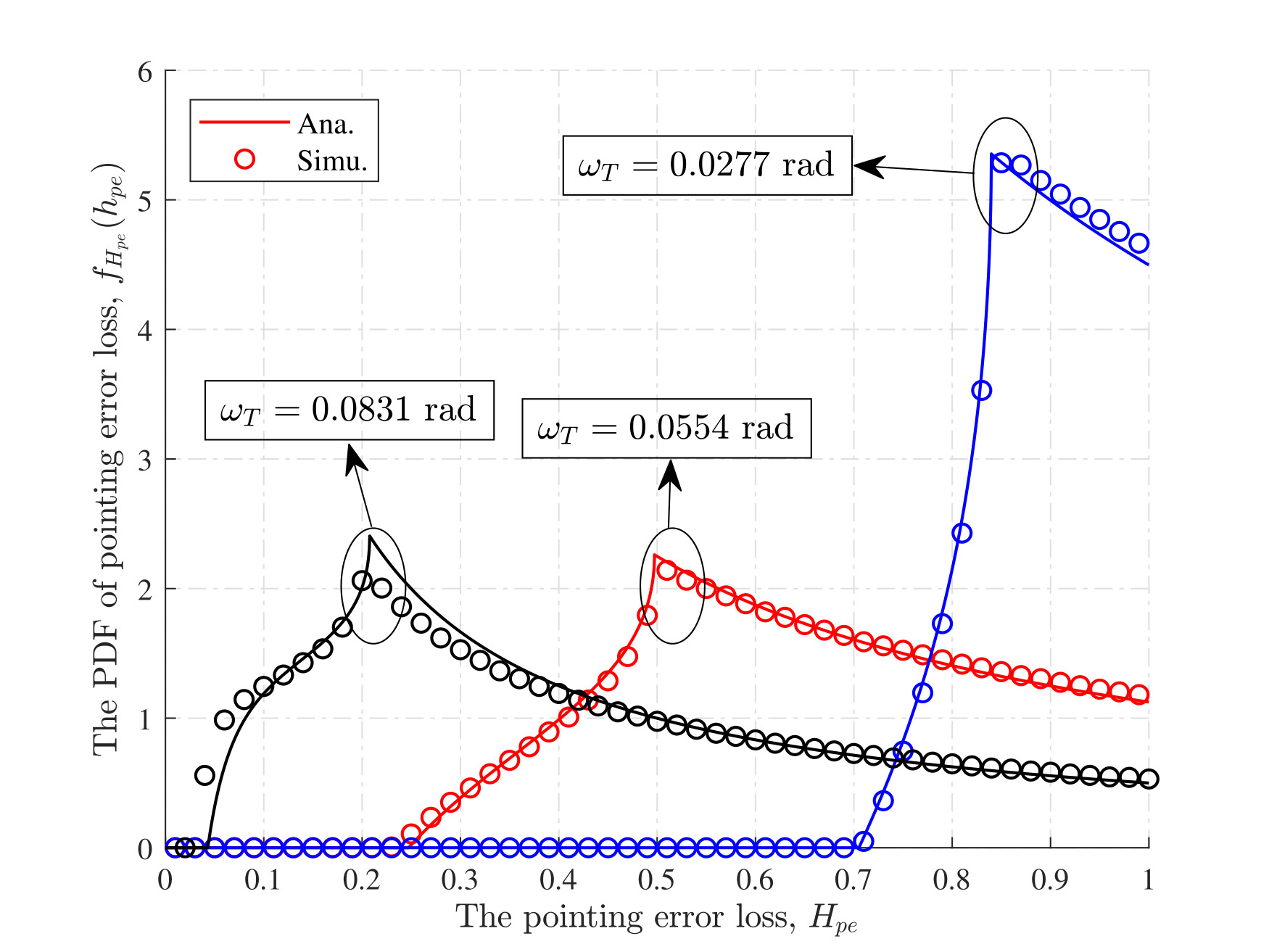}
    \vspace{-1em}
    \caption{The PDF of the pointing error loss $f_{H_{pe}}(h_{pe})$.}
    \vspace{-1.5em}
    \label{fig:Hpe}
\end{figure}

Fig.~\ref{fig:Hpe} plots the PDF of the pointing error loss, $f_{H_{pe}}(h_{pe})$. We first observe that our analytical results in Lemma~\ref{Lemma:1} align well with the simulation results, validating the correctness of our analysis. We then observe that $f_{H_{pe}}(h_{pe})$ first increases and then gradually decreases as $h_{pe}$ increases. This observation is due to the fact that moderate pointing errors occur with higher probability than either near-perfect beam alignment or severe beam misalignment, leading to a concentration of probability mass at intermediate pointing loss values. In addition, we find that the beamwidth $\omega_T$ has a pronounced impact on the pointing error loss distribution. Specifically, a smaller $\omega_T$ reduces the severity of pointing errors and improves the effective link quality. However, reducing $\omega_T$ generally requires finer beam resolution and more accurate beam training, which may increase training overhead and system complexity. This result highlights the importance of carefully selecting the beamwidth in indoor THz systems to balance alignment accuracy and beam training efficiency.


\begin{figure}[t]
    \centering
        \subfigure[At Location 2.]{
            \includegraphics[width=0.9\columnwidth]{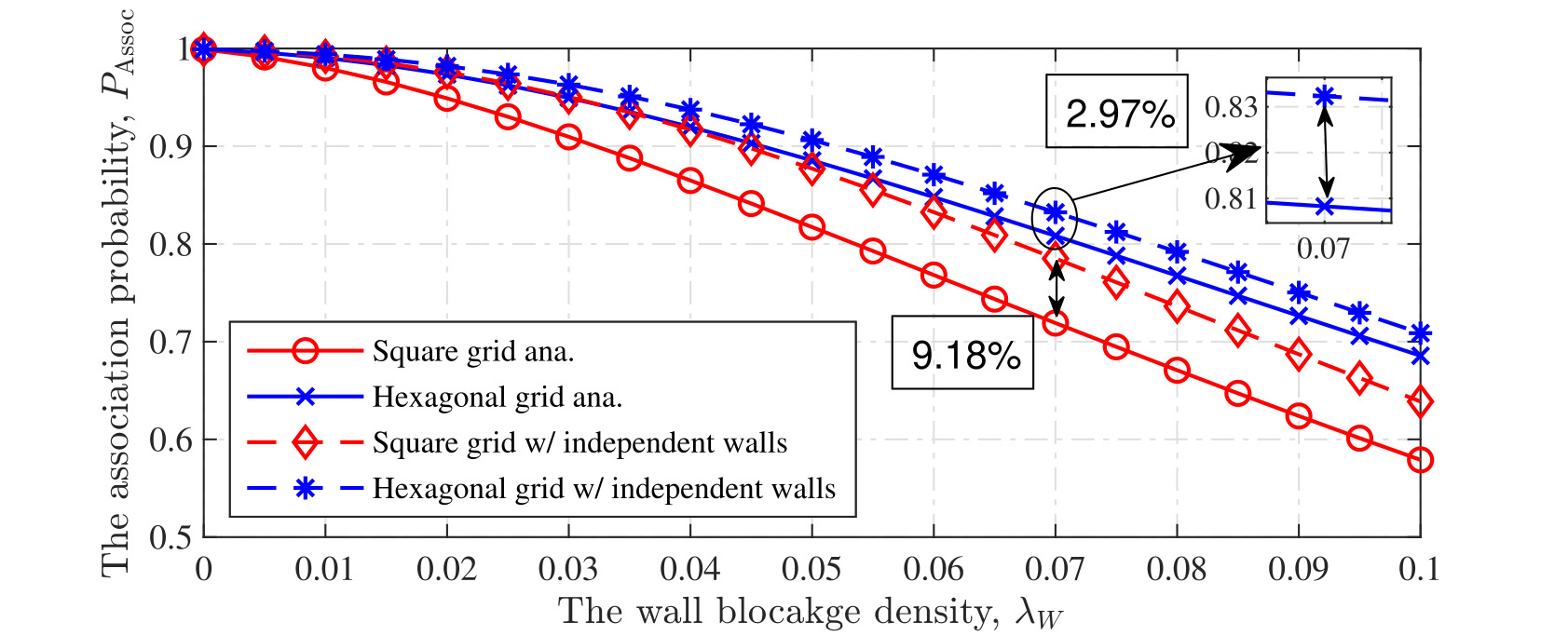}
            }\label{fig:AssocP2}\vspace{-0.5em}
        \subfigure[At Location 3.]{
            \includegraphics[width=0.9\columnwidth]{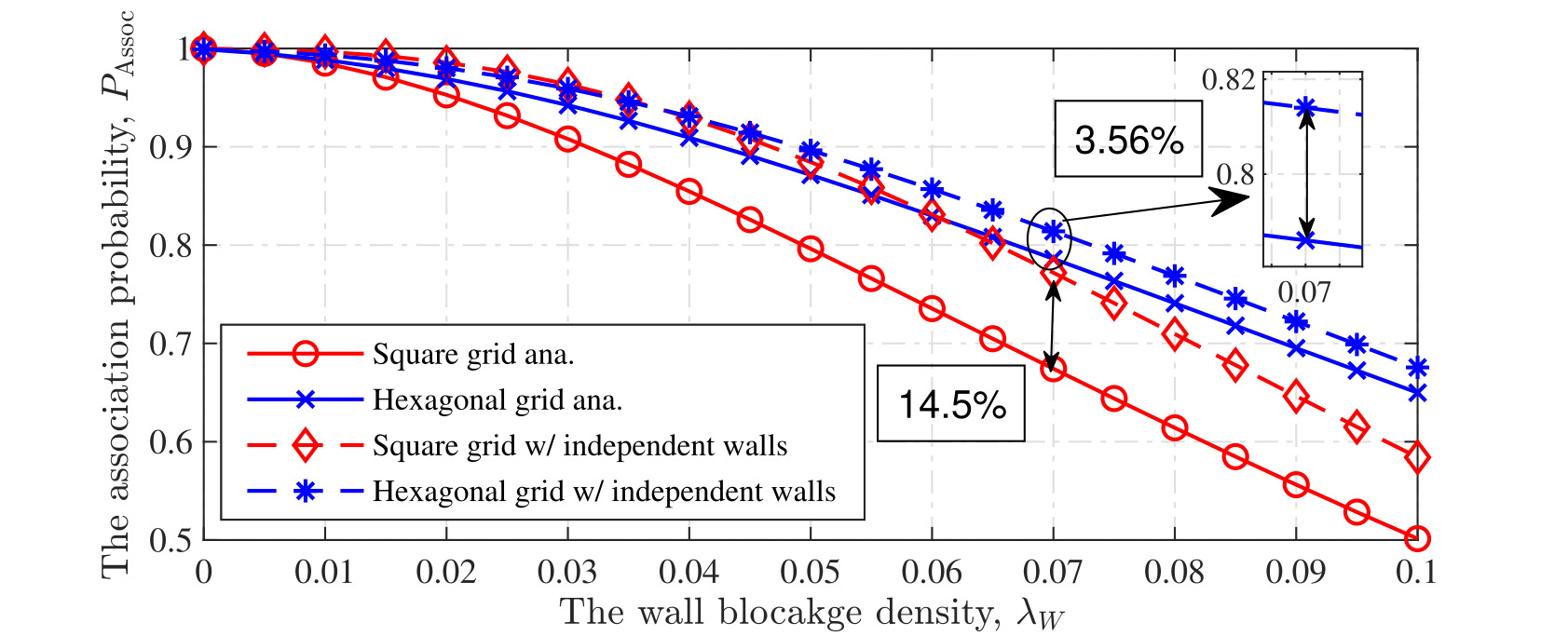}
            }\label{fig:AssocP3}\vspace{-0.5em}
        \caption{The association probability of $U_0$, $P_{\mathrm{Assoc}}$, versus the wall blockage density, $\lambda_W$, with different $U_0$ locations.}\vspace{-1.5em}
    \label{fig:AssocP}
\end{figure}

Fig.~\ref{fig:AssocP} plots the association probability, $P_{\mathrm{Assoc}}$, versus the wall blockage density, $\lambda_{w}$. We first observe that $P_{\mathrm{Assoc}}$ monotonically decreases as $\lambda_{w}$ increases, since a higher wall density increases the likelihood of link blockages and thus reduces the probability that a UE can establish a LoS association with an AP. We then observe that, both grid structures, $P_{\mathrm{Assoc}}$ is smaller that obtained under the assumption of independent wall blockages across different APs \cite{Wu2021TWC,Shafie2021JSAC,Kouzayha2023twc}. This discrepancy indicates that neglecting blockage correlation leads to an overestimation of the association probability. Therefore, blockage correlation should not be ignored in accurate performance analysis, which is consistent with \textit{Remark 1} and underscores the importance of our analytical framework. By further comparing the two deployment strategies, we observe that the hexagonal grid AP deployment achieves a higher association probability than the square grid AP deployment, especially when $\lambda_W$ is large, for both $U_0$ locations. This performance gain can be attributed to the reduced spatial correlation of wall blockages across multiple AP–UE links in the hexagonal layout, together with the shorter average distances between the UE and its nearest AP, which jointly enhance the likelihood of maintaining LoS connectivity. Specifically, the hexagonal grid provides more uniform angular separation and spatial diversity among neighboring APs, such that dominant AP–UE links are less likely to traverse the same wall blockages. These effects jointly contribute to the higher association probability observed under dense wall blockage conditions. 

\begin{figure}[t]
    \centering
    \includegraphics[width=0.9\columnwidth]{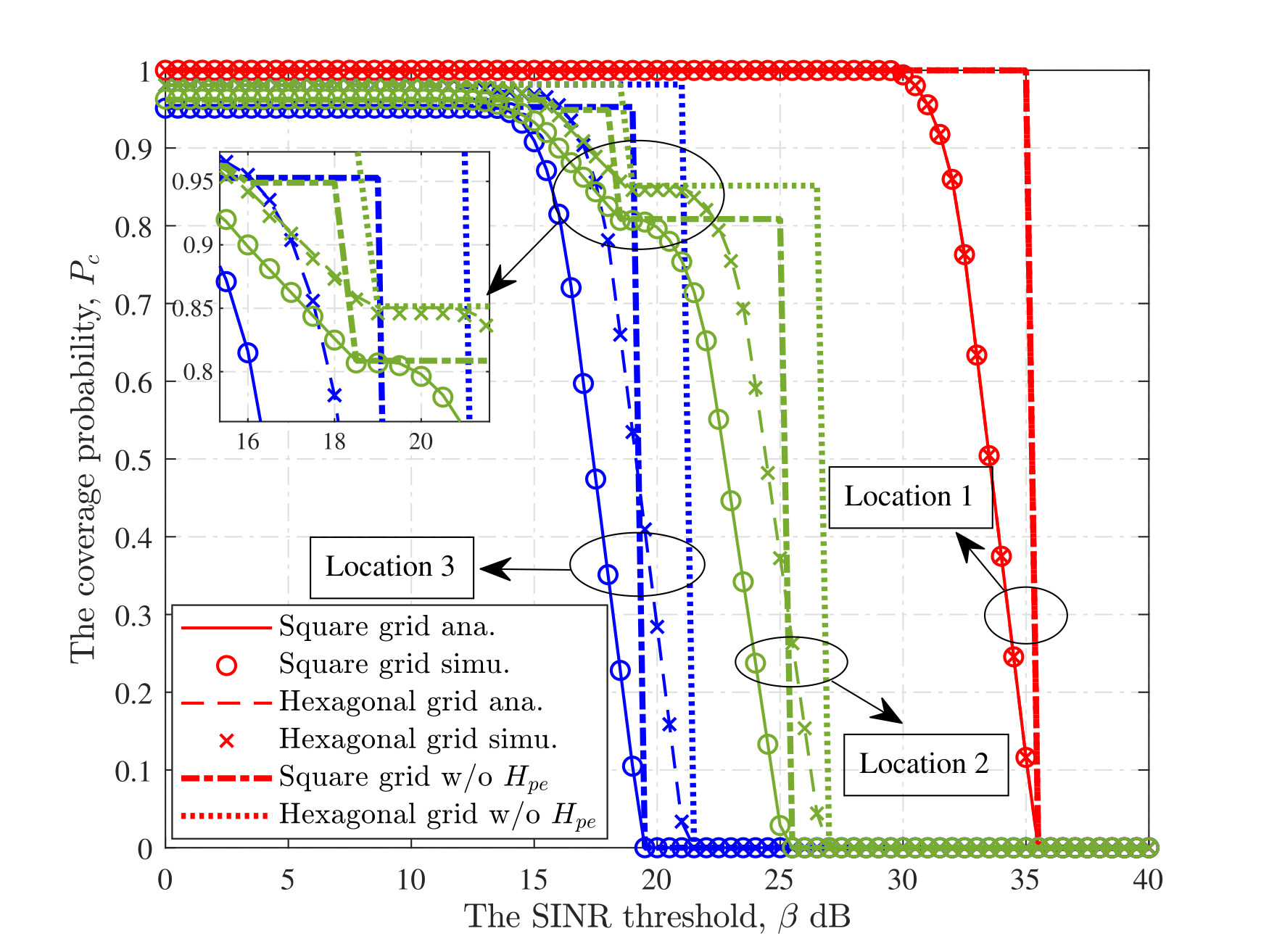}
    \vspace{-1em}
    \caption{The coverage probability of $U_0$, $P_c$, versus the $\mathrm{SINR}$ threshold, $\beta$ dB.}\vspace{-1.5em}
    \label{fig:CoverP}
\end{figure}

Fig.~\ref{fig:CoverP} plots the coverage probability of $U_0$, $P_c$, versus the $\mathrm{SINR}$ threshold, $\beta$. We first observe that our analytical results in Theorem~\ref{Theorem:Coverage} tightly match the simulation results, validating the accuracy of our analysis. We then observe that when $U_0$ is located directly beneath $\mathrm{AP}_{0,0}$ the coverage probabilities of the square and hexagonal grid AP deployments nearly coincide. In this case, the received signal is dominated by the serving AP, and the influence of the AP layout on the resulting SINR is marginal. As $U_0$ moves away from $\mathrm{AP}_{0,0}$, the coverage probability decreases, with the most pronounced degradation observed at Location~3, which corresponds to the farthest position from the nearest AP of $U_0$. This trend reflects the combined effects of increased path loss on the desired link and stronger interference from neighboring APs. Moreover, we observe that the hexagonal grid AP deployment consistently achieves higher coverage probability than the square grid AP deployment. This performance advantage becomes more evident at larger SINR thresholds, where links with shorter distances play a more critical role in maintaining coverage. We further observe a non-monotonic behavior when $U_0$ is located at Location~2, where the coverage probability first decreases rapidly, then remains relatively flat, and finally decreases again as the SINR threshold increases. The initial sharp decrease arises because, at moderate SINR thresholds, the received signals from non-nearest APs are generally insufficient to satisfy the SINR requirement. Consequently, enlarging the coverage boundary $R$ does not improve the coverage probability under relatively high SINR requirements, as the performance is primarily limited by the quality of the desired link rather than the availability of additional APs. As the SINR threshold increases further, the coverage probability continues to decrease due to the increasingly stringent SINR requirement, under which even the desired link from the serving AP cannot reliably sustain the required SINR due to the combined effects of path loss, residual pointing error, interference, and noise. Finally, the impact of pointing errors is consistently observed across all UE locations. When pointing errors are taken into account, the coverage probability is reduced over the entire range of SINR thresholds, with the corresponding curves lying below those obtained under perfect beam alignment. This degradation is attributed to the additional beamforming loss caused by angular misalignment, which reduces the effective array gain of the desired link. Moreover, the impact of pointing errors becomes more pronounced as $U_0$ moves farther away from its nearest AP, since longer link distances and narrower beams make the received signal more sensitive to alignment inaccuracies.

\begin{figure}[t]
    \centering
    \includegraphics[width=0.9\columnwidth]{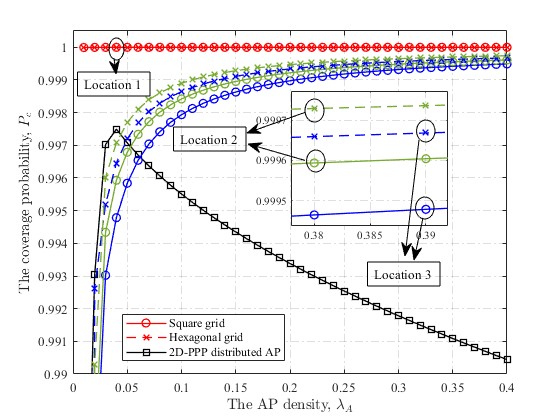}
    \vspace{-1em}
    \caption{The coverage probability of $U_0$, $P_c$, versus the AP density, $\lambda_A$, with $\beta=20$ dB.}
    \vspace{-1.5em}
    \label{fig:CoverPvsLamb}
\end{figure}

Fig.~\ref{fig:CoverPvsLamb} plots the coverage probability of the typical UE $U_0$, denoted by $P_c$, versus the AP density $\lambda_A$. The average inter-AP distance is inversely related to the AP density and is given by $d_{\mathrm{AP}}=\sqrt{1/(c_2\lambda_A)}$. We first observe that, for both square and hexagonal grid AP deployments, the coverage probability monotonically increases with $\lambda_A$ at Location~2 and Location~3. This is because a higher AP density shortens the distance between the UE and its serving AP, thereby strengthening the desired signal power. Meanwhile, the resulting increase in interference remains relatively moderate due to the use of directional antennas and the presence of blockages, leading to a net improvement in the received SINR. Moreover, for a given AP density, the hexagonal grid AP deployment consistently achieves a higher coverage probability than the square grid AP deployment, further demonstrating the performance advantage of hexagonal deployments. In contrast, when APs are distributed according to a 2D-PPP, the coverage probability first increases and then decreases as $\lambda_A$ increases. The initial increase is again attributed to the reduced UE–serving AP distance. However, as $\lambda_A$ continues to grow, the interference power rises more rapidly due to randomly located nearby interferers, which eventually outweighs the signal power gain and leads to a degradation in coverage probability. Moreover, by comparing structured grid AP deployments with the PPP-based AP deployment, we find that grid-based layouts significantly outperform the PPP case, particularly at moderate-to-high AP densities, which is consistent with the observations reported in \cite{Andrews2011tcom}. This performance gap arises because structured deployments offer more regular UE–AP distances and improved control over interference and blockage correlation, whereas random deployments are more susceptible to strong nearby interferers that degrade the SINR. These results further suggest that while stochastic geometry–based models provide useful insights, they may not fully capture the performance characteristics of indoor THz systems under practical, structured deployment scenarios.

\begin{figure}[t]
    \centering
    \includegraphics[width=0.9\columnwidth]{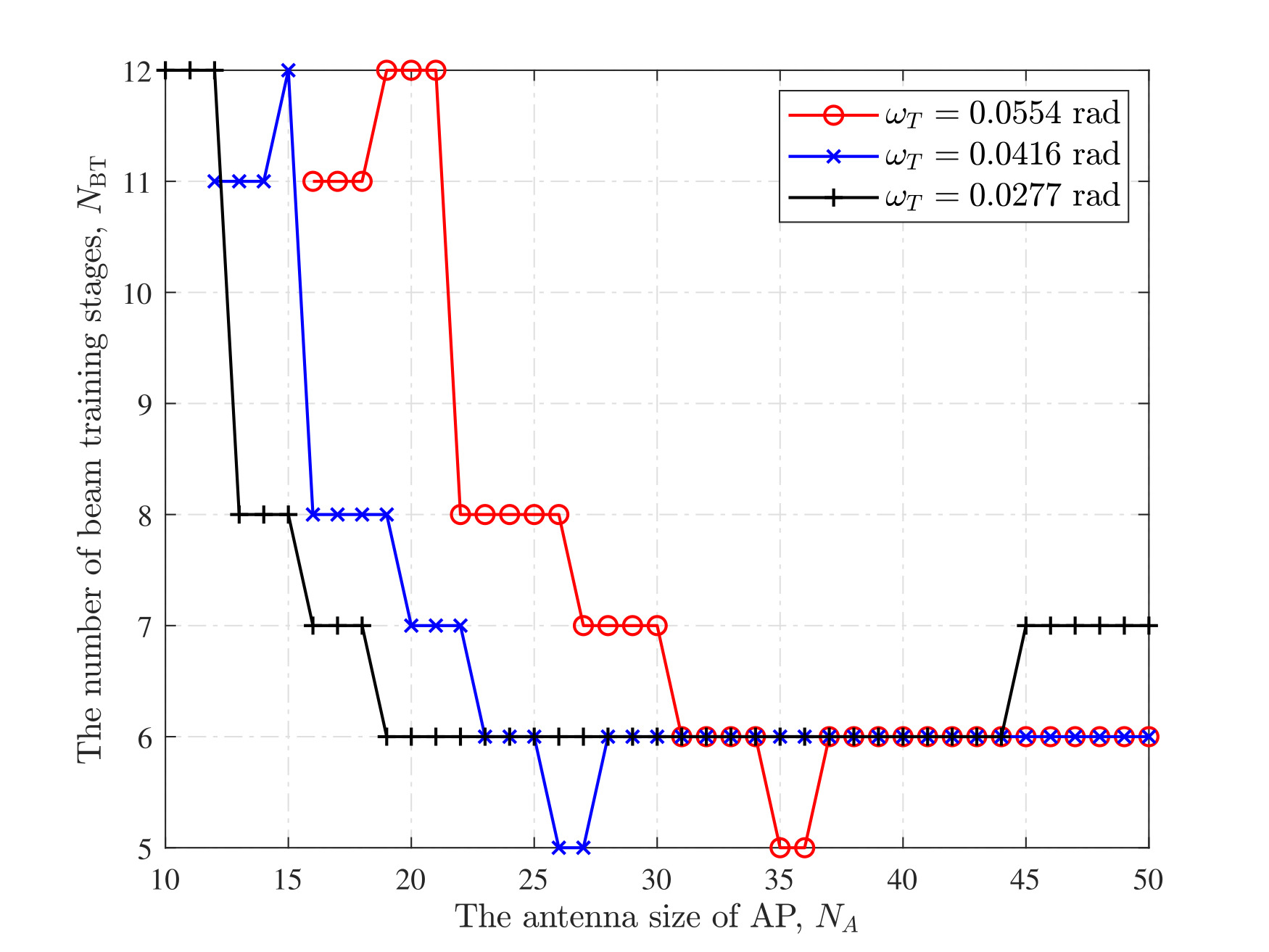}
    \vspace{-1em}
    \caption{The number of beam training stages, $N_{\mathrm{BT}}$, versus the antenna size of AP, $N_A$.}
    \vspace{-1.5em}
    \label{fig:Beamtrain}
\end{figure}

Fig.~\ref{fig:Beamtrain} plots the number of beam training stages, $N_{\mathrm{BT}}$, versus the antenna size of AP, $N_A$. We first observe that $N_{\mathrm{BT}}$ decreases initially and then slightly increases as $N_A$ increases. This observation can be explained by the competing effects identified in \textit{Remark 2}. Specifically, increasing in the antenna array size enhances the antenna gain and training parallelism, thereby reducing the required number of stages. However, once the RF-chain constraint is reached, further enlarging the array narrows the beamwidth and increases the number of candidate beams, which slightly increases $N_{\mathrm{BT}}$. 
We further observe that smaller beamwidths $\omega_T$ lead to fewer training stages when $N_A$ is small, as narrower beams provide higher antenna gain and improve the beam training SINR. In contrast, for large $N_A$, smaller $\omega_T$ can increase $N_{\mathrm{BT}}$ due to the larger number of beams required to cover the angular domain. These results reveal a fundamental trade-off among antenna size, beamwidth selection, and beam training efficiency, highlighting the importance of jointly optimizing the antenna array and beamwidth in THz system design.

\section{Conclusion}\label{Sec:Conclusion}

In this paper, we investigated the coverage probability of downlink transmission in a 3D indoor THz communication system by jointly considering structured AP deployment topology, wall blockage correlation, beam training, and pointing error. A comprehensive analytical framework was developed to characterize the impact of practical indoor propagation and deployment features on system performance. The analytical and numerical results demonstrated that wall blockage correlation across multiple APs significantly degrades association and coverage probabilities, and its impact cannot be neglected in indoor THz system analysis. Compared with square deployments, the hexagonal grid AP deployment consistently achieves superior performance by mitigating correlated blockage effects and reducing the distance between UEs and their nearest APs. In addition, coverage performance was shown to strongly depend on UE location, with noticeable degradation observed as the UE moves away from the serving AP due to increased path loss and stronger interference. The results further revealed that residual pointing error introduces substantial coverage loss across all UE locations, with its impact becoming more pronounced for longer links and narrower beams. Moreover, increasing the antenna array size does not necessarily reduce the overall beam training overhead, since the benefits of higher antenna gain may be offset by the increased number of beams associated with narrower beamwidths and higher sensitivity to beam misalignment. These findings suggest that future indoor THz systems should adopt joint optimization of deployment topology, beam training strategy, and antenna configuration, rather than optimizing each component in isolation.

\begin{appendices}

\vspace{-0.5em}
\section{Proof of Lemma~\ref{Lemma:1}}\label{Appendix:A}

Since the UE is assumed to be uniformly distributed within the coverage region of a training beam, the corresponding azimuth and elevation angular offsets observed from the AP can be modeled as independent and identically distributed random variables, i.e., $\theta_{A,H}$ and $\theta_{A,V} \sim \mathrm{Unif}(-\omega_T,\omega_T)$. As defined in the system model, for small angular deviations, the squared angular offset at the AP is given by $\theta_A^2 = \theta_{A,H}^2 + \theta_{A,V}^2$. Thus, the PDF of $\theta_A^2$ is calculated as
\begin{align}\label{eq:fTheta}
f_{\theta_A^2}(\Theta)
&=\frac{1}{4\omega_{T}^2}\int_{\psi_1}^{\psi_2}\frac{1}{\sqrt{\psi(\Theta-\psi)}}\mathrm{d}\psi,
\end{align}
where $\psi_1=\max\{0,\Theta-\omega_T^2\}$ and $\psi_2=\min\{\omega_T^2,\Theta\}$ for $0<\Theta<2\omega_T^2$. By substituting $\psi = \Theta\sin^2\phi$ into \eqref{eq:fTheta}, where $\phi \in [0,\tfrac{\pi}{2}]$, we obtain
\begin{align}
f_{\theta_A^2}(\Theta) &=\frac{1}{4\omega_{T}^2}\int_{\psi_1}^{\psi_2}\frac{1}{\Theta\sin\phi\cos\phi}\mathrm{d} \Theta\sin^2\phi\notag\\
&= \frac{1}{2\omega_{T}^2}(\psi_4 - \psi_3),
\end{align}
where $\psi_3 = \arcsin\sqrt{\psi_1/\Theta}$ and $\psi_4 = \arcsin\sqrt{\psi_2/\Theta}$. The PDF of $\theta_A^2$ is obtained by separately considering two cases.  
For $0<\Theta<\omega_T^2$, we have $\psi_1=0$ and $\psi_2=\Theta$, which yields $\psi_3=0$ and $\psi_4=\pi/2$, and hence
\begin{align}\label{eq:fthetaA21}
f_{\theta_A^2}(\Theta) = \frac{\pi}{4\omega_{T}^2}.
\end{align}
For $\omega_T^2\le \Theta<2\omega_T^2$, the limits become $\psi_1=\Theta-\omega_T^2$ and $\psi_2=\omega_T^2$, leading to
$\psi_4=\arcsin\!\left(\omega_T/\sqrt{\Theta}\right)$ and
$\psi_3=\frac{\pi}{2}-\arcsin\!\left(\omega_T/\sqrt{\Theta}\right)$.
Accordingly, the PDF is given by
\begin{align}\label{eq:fthetaA22}
f_{\theta_A^2}(\Theta)
= \frac{1}{\omega_{T}^2}\left(\arcsin\!\left(\frac{\omega_{T}}{\sqrt{\Theta}}\right) - \frac{\pi}{4}\right).
\end{align}
Combining the \eqref{eq:fthetaA21} and \eqref{eq:fthetaA22}, the PDF of $\theta_A^2$ can be summarized as
\begin{align}
\label{eq:f_theta2}
f_{\theta_A^2}(\Theta)\! =\!\left\{
\begin{aligned}
&\frac{\pi}{4\omega_{T}^2}, &&\textrm{if }0 \leq \Theta \leq \omega_{T}^2,\\
&\frac{1}{\omega_{T}^2}\!\left(\arcsin\!\frac{\omega_{T}}{\sqrt{\Theta}}\! -\! \frac{\pi}{4}\right), &&\textrm{if } \omega_{T}^2 < \Theta \leq 2\omega_{T}^2,\\
&0, && \text{otherwise.}
\end{aligned}\right.
\end{align}
Accordingly, the main-lobe gain of the AP antenna can be well approximated by a Gaussian beam pattern, expressed as  
\begin{align}
    H_{pe}(\theta_A)\approx \exp\left(-\theta_A^2/\omega_A^2\right).
\end{align}
Since $H_{pe}(\theta_A)$ is a monotone decreasing function of $\theta_A^2$, with inverse $\theta_A^2=-\omega_A^2\ln H_{pe}$, the PDF of $H_{pe}$ follows from the change-of-variables theorem as
\begin{align}\label{eq:fHTheta}
f_{H_{pe}}(h) = f_{\theta_A^2}(\Theta) \, \left|\frac{d\Theta}{dh}\right|
= f_{\theta_A^2}(-\omega_A^2 \ln h) \cdot \frac{\omega_A^2}{h},
\end{align}
for $h\in\big(e^{-2\omega_T^2/\omega_A^2},\,1\big]$. Substituting \eqref{eq:f_theta2} into \eqref{eq:fHTheta} yields the final PDF expression in \eqref{eq:fhpe}. The corresponding CDF follows directly from
\begin{align}
F_{H_{pe}}(h)=\int_0^{h}f_{H_{pe}}(h_1)\,\mathrm{d}h_1,
\end{align}
which leads to \eqref{eq:Fchpe}.
\vspace{-0.5em}
\section{Proof of Lemma~\ref{Lemma:3}}\label{Appendix:B}
We first show that the aggregate interference is finite. Here, we define an upper-bounding set of interfering APs as $\Psi_{\mathrm{AP}_{i,j}}^{\mathrm{up}} = \{\mathrm{AP}_{m,n}||m|\geq|i| \textrm{ or } |n|\geq|j|\}$, which satisfies $\Psi_{\mathrm{AP}_{i,j}} \subseteq \Psi_{\mathrm{AP}_{i,j}}^{\mathrm{up}}$. Due to the symmetry of the grid AP deployment and the monotonicity of the path loss function, for any $m,n \ge 0$, we have $W(d_{m,n}) \ge W(d_{\pm m,\pm n})$. Hence, the aggregate interference can be upper-bounded as
    \begin{align}\label{eq:UpperIntf}
        I_{i,j}\leq 4 P_t G_{s} \xi &\Bigg(\sum_{m=0}^{\infty}\sum_{n=0}^{\infty} p_{m,n}W(d_{m,n}) \notag\\
        &- \sum_{m=0}^{|i|-1}\sum_{n=0}^{|j|-1}p_{m,n}W(d_{m,n})\Bigg).
    \end{align}
The second term in \eqref{eq:UpperIntf} involves a finite number of APs and is therefore finite. It thus suffices to establish the convergence of the first double summation. Recalling that $p_{m,n} = \exp(-\chi_{m,n})$, we have
\begin{align}
p_{m,n}W(d_{m,n})
= \frac{e^{-\chi_{m,n}}}{d_{m,n}^2+\Delta h^2}
  e^{-\epsilon(f)\sqrt{d_{m,n}^2+\Delta h^2}}.
\end{align}
Since $\sqrt{d_{m,n}^2+\Delta h^2} \ge d_{m,n}$, it follows that
\begin{align}
p_{m,n}W(d_{m,n})
&\leq \frac{e^{-(\chi_{m,n}+\epsilon(f)d_{m,n})}}{d_{m,n}^2+\Delta h^2}.
\end{align}
Therefore, it is sufficient to prove the convergence of
\begin{align}
S = \sum_{m,n} \frac{e^{-(\alpha+\lambda_W+\epsilon(f)) d_{m,n}}}{d_{m,n}^2+\Delta h^2}.
\end{align}
For both square and hexagonal grid AP deployments, the inter-AP distance satisfies
\begin{align}
    \frac{d_{m,n}}{d_{\mathrm{AP}}}
\geq
c_2 \sqrt{(m-x_0)^2 + (n-y_0)^2}
\triangleq c_2 r_{m,n},
\end{align}
which leads to
\begin{align}\label{eq:Supperboun1}
S 
&\leq \frac{1}{c_2^2}\sum_{m,n}\frac{e^{-\gamma r_{m,n}}}{r_{m,n}^2+\delta^2},
\end{align}
where $\gamma = (\alpha + \lambda_W + \epsilon(f)) c_2$ and $\delta = \Delta h / c_2$. Next, we partition the lattice into annuli centered at the UE location and define $\aleph_k = \{(m,n): k \leq r_{m,n} < k+1,\ k\in\mathbb{N}\}$. 
For both grid AP deployments, the number of APs within radius $R$ is upper-bounded by $\pi R^2 / d_{\mathrm{AP}}^2$, implying $|\aleph_k| \le \frac{\pi}{d_{\mathrm{AP}}^2} k$ for all $k \ge 1$. For any $(m,n)\in \aleph_k$, we have
\begin{align}
\frac{e^{-\gamma r_{m,n}}}{r_{m,n}^2+\delta^2} 
\le \frac{e^{-\gamma k}}{k^2}.
\end{align}
Hence, \eqref{eq:Supperboun1} can be further upper-bounded by summing over all annuli as
\begin{align}
S
&\le \frac{1}{c_2^2}
\sum_{k=1}^{\infty}
\sum_{(m,n)\in \aleph_k}
\frac{e^{-\gamma r_{m,n}}}{r_{m,n}^2+\delta^2}\le \frac{1}{c_2^2}
\sum_{k=1}^{\infty}
|\aleph_k|\,
\frac{e^{-\gamma k}}{k^2} \notag\\
&\le
\frac{\pi}{c_2^2 d_{\mathrm{AP}}^2}
\sum_{k=1}^{\infty}
\frac{k e^{-\gamma k}}{k^2} =
\frac{\pi}{c_2^2 d_{\mathrm{AP}}^2}
\sum_{k=1}^{\infty}
\frac{e^{-\gamma k}}{k}.
\end{align}
The series $\sum_{k=1}^{\infty} \exp(-\gamma k)/k$ converges for all $\gamma>0$, since
\begin{align}
\sum_{k=1}^{\infty} \frac{e^{-\gamma k}}{k}
\le \int_{0}^{\infty} \frac{e^{-\gamma x}}{x+1}\,dx 
= E_1(\gamma) < \infty,
\end{align}
where $E_1(\cdot)$ denotes the exponential integral function. Therefore, $S$ is finite, which establishes the convergence of the aggregate interference.


Next, the mean interference conditioned on $\mathcal{A}_{i,j}$ is given by
        \begin{align}
            \mu_{I_{i,j}} &= \E[I_{i,j}] =  \sum\limits_{(m,n)\in \Psi_{i,j}  }\E[P_{m,n}|\mathcal{A}_{i,j}],
        \end{align}
which yields \eqref{eq:averageI}. Similarly, the variance of the interference can be expressed as
        \begin{align}
            &\sigma_{I_{i,j}}^2 = \mathrm{Var}\left(I_{i,j}\right) = \sum_{(m,n)\in \Psi_{i,j}}  \mathrm{Var}\left(P_{m,n}\big|\overline{\mathcal{B}}_{i,j}^W=1\right) \notag\\
        &+ \sum_{\substack{(m_1,n_1)\!\neq\! (m_2,n_2)}} \!\mathrm{Cov}\left(P_{m_1,n_1},P_{m_2,n_2}\big|\overline{\mathcal{B}}_{i,j}^W=1\right), 
        \end{align}
which leads to \eqref{eq:varianceI}.

\section{Proof of Lemma~\ref{lem:PA_exact}}\label{Appendix:Lemma4}
The association event $\mathcal{A}_{i,j}$ occurs if $\mathrm{AP}_{i,j}$ is unblocked by both human and wall blockages, and all APs in $\overline{\Psi}_{i,j}$ fail to provide a LoS link due to either human or wall blockage. Accordingly, the association probability can be calculated by
\begin{align}
\Pr(\mathcal{A}_{i,j}) &\! =\! \E_{\mathcal{B}^W}\left [p^H_{i,j} \overline{\mathcal{B}}_{i,j}^W \prod_{(m,n)\in \overline{\Psi}_{i,j}} \big(1 - p^H_{m,n} \, \overline{\mathcal{B}}_{m,n}^W \big) \right]\notag\\
&\!=\! \E_{\mathcal{B}^W}\left [\!\sum_{\Phi \subseteq \overline{\Psi}_{i,j}} (-1)^{|\Phi|} 
\prod_{(m,n) \in \Phi} p^H_{m,n} \, \overline{\mathcal{B}}_{m,n}^W\right],
\end{align}
where $\overline{\mathcal{B}}_{m,n}^W = 1-{\mathcal{B}}_{m,n}^W$. By taking the expectation over the wall blockage process, we obtain the joint probability that the considered APs are unblocked by walls, obtained as
\begin{align}
&\E_{\mathcal{B}^W}\left[\overline{\mathcal{B}}_{i,j}^W \prod_{(m,n) \in \Phi}\overline{\mathcal{B}}_{m,n}^W\right]\notag\\
&=\Pr\Big(\mathcal{B}_{i,j}^W = 0, \ \mathcal{B}_{m,n}^W = 0 \ \forall (m,n)\in \Phi\Big) = p^W_{(i,j) \cup \Phi},
\end{align} 
where $p^W_{(i,j) \cup \Phi}$ is the joint probability that all links from the user to $\mathrm{AP}_{i,j}$ and all APs in $\Phi$ are simultaneously unblocked by walls, obtained as
\begin{align}\label{eq:jointWub}
    p^W_{(i,j) \cup \Phi} = \exp(-\lambda_W(U_{(i,j),\Phi,x}+U_{(i,j),\Phi,y})).
\end{align}
Thus, the association probability is calculated as
\begin{align}\label{eq:PrAij1}
\Pr(\mathcal{A}_{i,j}) 
\!=\! p^H_{i,j}\! \sum_{\Phi \subseteq \overline{\Psi}_{i,j}}\!(\!-\!1)^{|\Phi|} 
\left( \prod_{(m,n) \in \Phi} p^H_{m,n} \right) p^W_{(i,j) \cup \Phi}.
\end{align}
By substituting \eqref{eq:jointWub} into \eqref{eq:PrAij1}, we obtain \eqref{eq:PrAij}.

\section{Proof of Theorem~\ref{Theorem:Coverage}}\label{Appendix:C}
Given the association event $\mathcal{A}_{i,j}$, the conditional coverage probability
$\Pr(\mathrm{SINR}>\beta \mid \mathcal{A}_{i,j})$ can be expressed as
\begin{align}\label{eq:PcoverAij}
\Pr(\mathrm{SINR}>\beta \mid &\mathcal{A}_{i,j})
=
\Pr\!\left(
H_{pe} >
\frac{(I_{i,j}+N_0)\beta}{\zeta_{i,j}}
\Bigg| \mathcal{A}_{i,j}
\right) \notag\\
&=
\mathbb{E}_{I_{i,j}}
\left[
1 - F_{H_{pe}}\!\left(
\frac{(I_{i,j}+N_0)\beta}{\zeta_{i,j}}
\right)
\right] \notag\\
&=
1-\mathbb{E}_{I_{i,j}}
\left[
F_{H_{pe}}\!\left(
\frac{(I_{i,j}+N_0)\beta}{\zeta_{i,j}}
\right)
\right].
\end{align}
The above expectation does not admit a tractable closed-form expression due to the complicated distribution of the aggregate interference $I_{i,j}$ and the piecewise nature of $F_{H_{pe}}(\cdot)$. To obtain a tractable approximation, we apply a second-order Taylor expansion. Here, we define $g(x) \triangleq F_{H_{pe}}\!\left(\frac{(x+N_0)\beta}{\zeta_{i,j}}\right)$. Expanding $g(I_{i,j})$ around $\mu_{I_{i,j}}=\mathbb{E}[I_{i,j}]$ yields
\begin{align}
\mathbb{E}[g(I_{i,j})]
&\approx
g(\mu_{I_{i,j}})
+\frac{g''(\mu_{I_{i,j}})}{2}\,\sigma_{I_{i,j}}^2.
\end{align}
Expanding $g(I_{i,j})$ around $\mu_{I_{i,j}}=\mathbb{E}[I_{i,j}]$ yields
\begin{align}
\mathbb{E}[g(I_{i,j})]
\approx
g(\mu_{I_{i,j}})
+\frac{g''(\mu_{I_{i,j}})}{2}\,\sigma_{I_{i,j}}^2.
\end{align}
Applying such expansions gives
\begin{align}\label{eq:EIijFHpe}
\mathbb{E}_{I_{i,j}}
&\left[
F_{H_{pe}}\!\left(
\frac{(I_{i,j}+N_0)\beta}{\zeta_{i,j}}
\right)
\right]
\approx\;
F_{H_{pe}}\!\left(
\frac{(\mu_{I_{i,j}}+N_0)\beta}{\zeta_{i,j}}
\right) \notag\\
&+
\frac{\beta^2\sigma_{I_{i,j}}^2}{2\zeta_{i,j}^2}\,
f'_{H_{pe}}\!\left(
\frac{(\mu_{I_{i,j}}+N_0)\beta}{\zeta_{i,j}}
\right).
\end{align}
Substituting \eqref{eq:EIijFHpe} into \eqref{eq:PcoverAij} and then substituting the result into \eqref{eq:PCovercal1}, together with \eqref{eq:averageI} and \eqref{eq:varianceI}, yields \eqref{eq:coverage}.

\section{Proof of Lemma~\ref{Lemma:5}}\label{Appendix:Lemma5}
To analyze the achievable number of channel-training beams $N_{\mathrm{ct}}$, we first evaluate the approximated interference based on \eqref{eq:averageI}. For a sufficiently dense lattice of APs with spacing $d_{\mathrm{AP}}$, $\mu_{I_R}$ can be approximated by a two-dimensional integral and change to polar coordinates $(r,\theta)$, with $r \geq R$ and $\theta \in [0,2\pi]$, given as

\begin{align}
&\mu_{I_{R}}=P_tG_{\mathrm{S}}\xi \sum_{d_{m,n} > R} p_{m,n}W(d_{m,n})\notag\\
&\approx  \frac{P_tG_{\mathrm{S}}\xi }{d_{AP}^2}\iint_{r>R} \exp(-\chi_{m,n}) W(r) \, \mathrm{d}^2 \mathbf{r}\notag\\
&\approx \frac{P_tG_{\mathrm{S}}\xi }{d_{AP}^2}\! \int_0^{2\pi}\!\!\int_{R}^{\infty}\! r e^{(-\lambda_W (r\!\sin\theta\!+\!r\!\cos\theta)\! -\!\alpha r)} W(r)\mathrm{d}r \, \mathrm{d}\theta.
\end{align}
We note that the integral can be approximated by noting $\sqrt{r^2 + \Delta h^2} \approx r$ for large $r$, which gives

\begin{align}\label{eq:muI1}
\mu_{I_{R}}
\approx \frac{P_tG^s\xi }{d_{AP}^2} \int_0^{2\pi}\int_{R}^{\infty} \frac{r e^{-\rho_3 r}}{r^2 + \Delta h^2}  \mathrm{d}r\mathrm{d}\theta,
\end{align}
where $\rho_3 = (\alpha + \epsilon(f)+\lambda_W (\sin\theta\!+\!\cos\theta))$. We evaluate the double integral in \eqref{eq:muI1} by first integrating over the radial distance $r$, which leads to 
\begin{align}
\int_{R}^{\infty} \frac{r \, e^{- \rho_3 r}}{r^2 + \Delta h^2}  \mathrm{d}r=& \frac{1}{2} \Big[ e^{i \rho_3 \Delta h} 
\mathrm{Ei}\big(- \rho_3 (R + i \Delta h) \big) \notag \\
&+ e^{-i \rho_3 \Delta h} 
\mathrm{Ei}\big(- \rho_3 (R - i \Delta h) \big) \Big],
\end{align}
where $\mathrm{Ei}(\cdot)$ is the exponential integral function. Since $R$ is much larger than $\Delta h$, we approximate $R^2 + \Delta h^2 \approx R^2$ and use the asymptotic expansion $\mathrm{Ei}(-x) \approx e^{-x}/x$, yielding a simplified approximated expression in \eqref{eq:Iapprox}. By substituting \eqref{eq:Iintra} and \eqref{eq:Iapprox} into \eqref{eq:SINRrequirement}, we obtain \eqref{eq:Nctmax}.
\end{appendices}

\bibliographystyle{IEEEtran} 
\bibliography{bibli}

\end{document}